\documentclass[12pt]{article}
\usepackage{amsmath}
\usepackage{graphicx}
\usepackage{natbib}

\usepackage{url} 

\newcommand{\blind}{0}

\usepackage[pdfpagemode={UseOutlines},bookmarks=true,bookmarksopen=true,
   bookmarksopenlevel=0,bookmarksnumbered=true,hypertexnames=false,
   colorlinks,linkcolor={blue},citecolor={blue},urlcolor={red},
   pdfstartview={FitV},unicode,breaklinks=true]{hyperref}
 
\usepackage{algorithm}
\usepackage{algorithmic}
\algsetup{linenosize=\small}
\usepackage{bm}
\usepackage{amsfonts}
\usepackage{caption}
\DeclareMathOperator*{\argmin}{argmin}

\usepackage{setspace}
\usepackage{optidef}
\usepackage{tabularx}
\usepackage{caption}
\usepackage{subcaption}
\usepackage[nameinlink]{cleveref}

\newcommand{\LP}{\mbox{\scriptsize LP}}
\newcommand{\GP}{\mbox{\scriptsize GP}}

\makeatletter
\DeclareFontFamily{U}{tipa}{}
\DeclareFontShape{U}{tipa}{m}{n}{<->tipa10}{}
\newcommand{\arc@char}{{\usefont{U}{tipa}{m}{n}\symbol{62}}}%
\newcommand{\arc}[1]{\mathpalette\arc@arc{#1}}
\newcommand{\arc@arc}[2]{%
  \sbox0{$\m@th#1#2$}%
  \vbox{
    \hbox{\resizebox{\wd0}{\height}{\arc@char}}
    \nointerlineskip
    \box0
  }%
}

\begin{document}

\sloppy 

\def\spacingset#1{\renewcommand{\baselinestretch}%
{#1}\small\normalsize} \spacingset{1}


\if0\blind
{
  \title{\bf Statistical analysis of locally parameterized shapes}
  \author{Mohsen Taheri\footnote{Mohsen Taheri, Department of Mathematics and Physics, University of Stavanger (UiS), Email: mohsen.taherishalmani@uis.no} \quad and \quad J\"orn Schulz\footnote{Assoc. Prof. J\"orn Schulz, Department of Mathematics and Physics, University of Stavanger (UiS), Email: jorn.schulz@uis.no} \\
    Department of Mathematics \& Physics, University of Stavanger}
  \maketitle
} \fi

\if1\blind
{
  \bigskip
  \bigskip
  \bigskip
  \begin{center}
    {\LARGE\bf Title}
\end{center}
  \medskip
} \fi

\bigskip
\begin{abstract}
The alignment of shapes has been a crucial step in statistical shape analysis, for example, in calculating mean shape, detecting locational differences between two shape populations, and classification. Procrustes alignment is the most commonly used method and state of the art. In this work, we uncover that alignment might seriously affect the statistical analysis. For example, alignment can induce false shape differences and lead to misleading results and interpretations. We propose a novel hierarchical shape parameterization based on local coordinate systems. The local parameterized shapes are translation and rotation invariant. Thus, the inherent alignment problems from the commonly used global coordinate system for shape representation can be avoided using this parameterization. The new parameterization is also superior for shape deformation and simulation. The method's power is demonstrated on the hypothesis testing of simulated data as well as the left hippocampi of patients with Parkinson's disease and controls.
\end{abstract}

\noindent%
{\it Keywords:} Local coordinate system, Parkinson's disease, Procrustes alignment, Skeletal representation, s-rep parameterization, s-rep hypothesis testing.
\vfill

\newpage
\spacingset{1.5} 

\section{Introduction}\label{sec:introduction}
\begin{figure}[ht]
     \centering
     \begin{subfigure}[b]{0.47\textwidth}
         \centering
         \boxed{\includegraphics[width=0.9\textwidth]{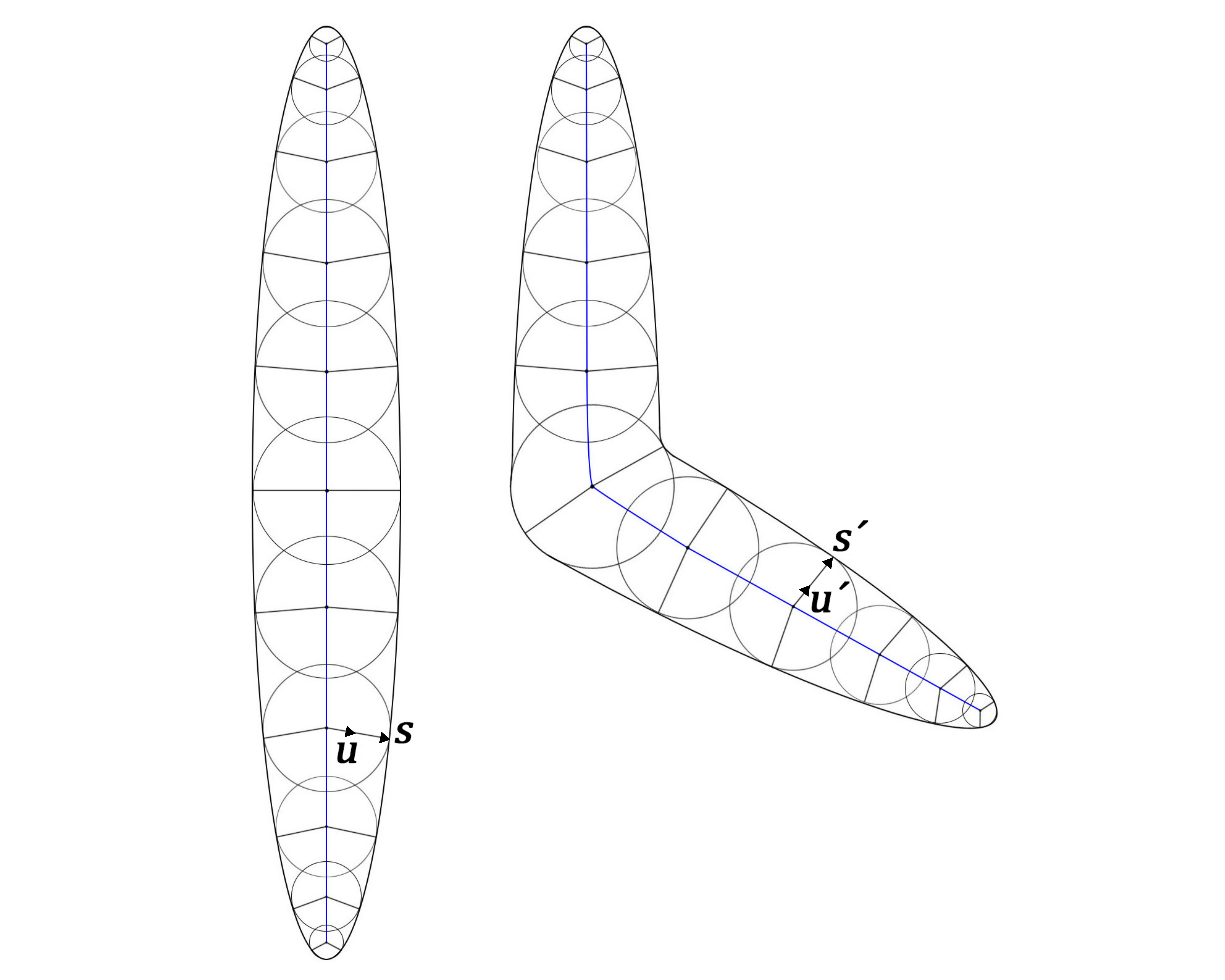}}
         \caption{\centering\footnotesize{m-rep}}
         \label{fig:m-rep1}
     \end{subfigure}
     \begin{subfigure}[b]{0.47\textwidth}
         \centering
         \boxed{\includegraphics[width=0.9\textwidth]{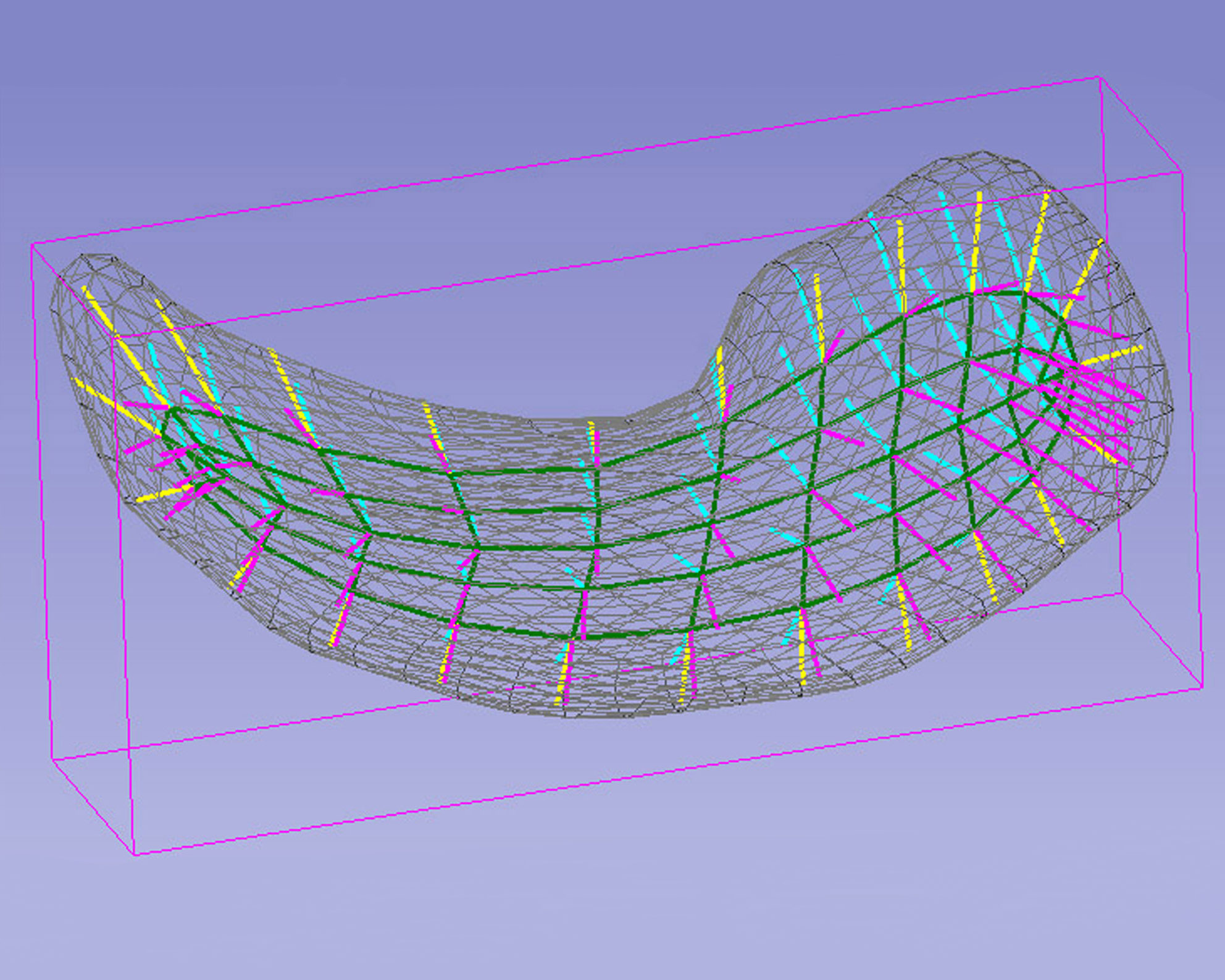}}
         \caption{\centering\footnotesize{ds-rep}}
         \label{fig:ds-rep1}
     \end{subfigure}
     \caption{\centering\footnotesize{Skeletal structure.
        (a) 2D m-reps of two ellipsoidal objects. $\bm{s}$ and $\bm{s}'$ are corresponding spokes with unit directions $\bm{u}$ and $\bm{u}'$. (b) A fitted ds-rep to a left hippocampus's mesh. Green, cyan, magenta, and yellow respectively indicate skeletal sheet, up spokes, down spokes, and crest spokes.}}
        \label{fig:skeletal_structure}
\end{figure}

In statistical shape analysis, besides classifying shapes, detecting and characterizing locational differences between two populations of shapes is a matter of special interest. Particularly in medical applications, shape analysis has the power to shed light on organ deformations, supporting diagnosis and treatment. Therefore, in a considerable number of articles, researchers try to find a set of corresponding \textit{geometric object properties} (GOPs) among shapes to explore regional dissimilarities. Corresponding GOPs can be represented in various ways, for example, by a set of landmarks on or inside the objects \citep[Ch.1]{dryden2016statistical}, a \textit{point distribution model} (PDM) such as \textit{spherical harmonics} PDM (SPHARM-PDM) \citep{Styner2006}, or as a subset of skeletal structures (e.g., a set of internal vectors) \citep[ch.8]{siddiqi2008medial}. For instance, \citep{cates2007shape} introduced entropy-based surface sampling PDM to compare brain objects of a group of patients with schizophrenia and a healthy \textit{control group} (CG). A large variety of studies can be found using PDM models to study human organs, e.g., \citep{alhadidi20123d, oguz2008cortical, achterberg2014hippocampal}. For the skeletal structure, \citep{pizer2013nested} introduced \textit{skeletal representation} (s-rep) and consequently \textit{discrete s-rep} (ds-rep) where ds-rep can be considered as a penalized version of \textit{medial representation} (m-rep) \citep{pizer1999segmentation} (see \Cref{fig:skeletal_structure}). An s-rep reflects the object's interior and defines a smooth implied boundary. \citep{schulz2016non} proposed a hypothesis test for ds-rep and studied the hippocampal differences between schizophrenia and CG.\par
Generally, these types of studies for shape analysis share three main steps. First, a preprocessing step, where both groups of objects are aligned based on their corresponding GOPs. The objective of the alignment is to quantify shape differences purely without locational information. Thus, the distance between objects is minimized to make them invariant under the act of Euclidean similarity transformations translation, rotation, and scaling. Usually, alignment takes place by \textit{generalized Procrustes analysis} (GPA) \citep[Ch.7]{dryden2016statistical}. Second, a set of partial hypothesis tests on GOPs to verify significant locational differences between the two groups. Third, multiple testing methods are applied to control false positives such as \textit{family wise error rate} (FWER), e.g., \citep{bonferroni1936teoria} or \textit{false discovery rate} (FDR), e.g., \citep{benjamini1995controlling}.\par
\begin{figure}[ht]
     \centering
     \begin{subfigure}[b]{0.327\textwidth}
         \centering
         \boxed{\includegraphics[width=0.95\textwidth]{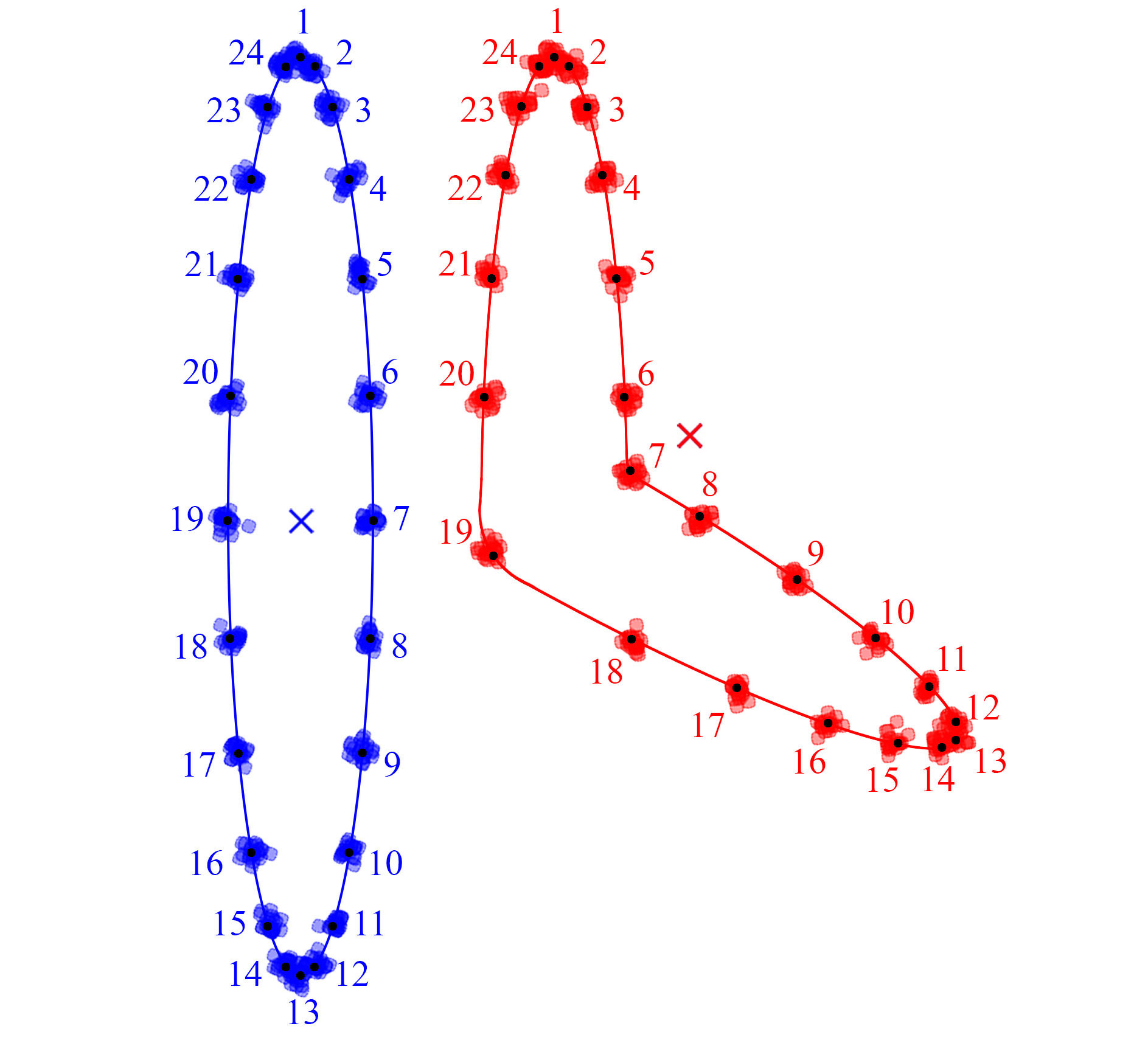}}
         \caption{\centering\footnotesize{Ellipsoidal PDMs}}
         \label{fig:arms_pdm_0}
     \end{subfigure}
     \hfill
     \begin{subfigure}[b]{0.327\textwidth}
         \centering
         \boxed{\includegraphics[width=0.95\textwidth]{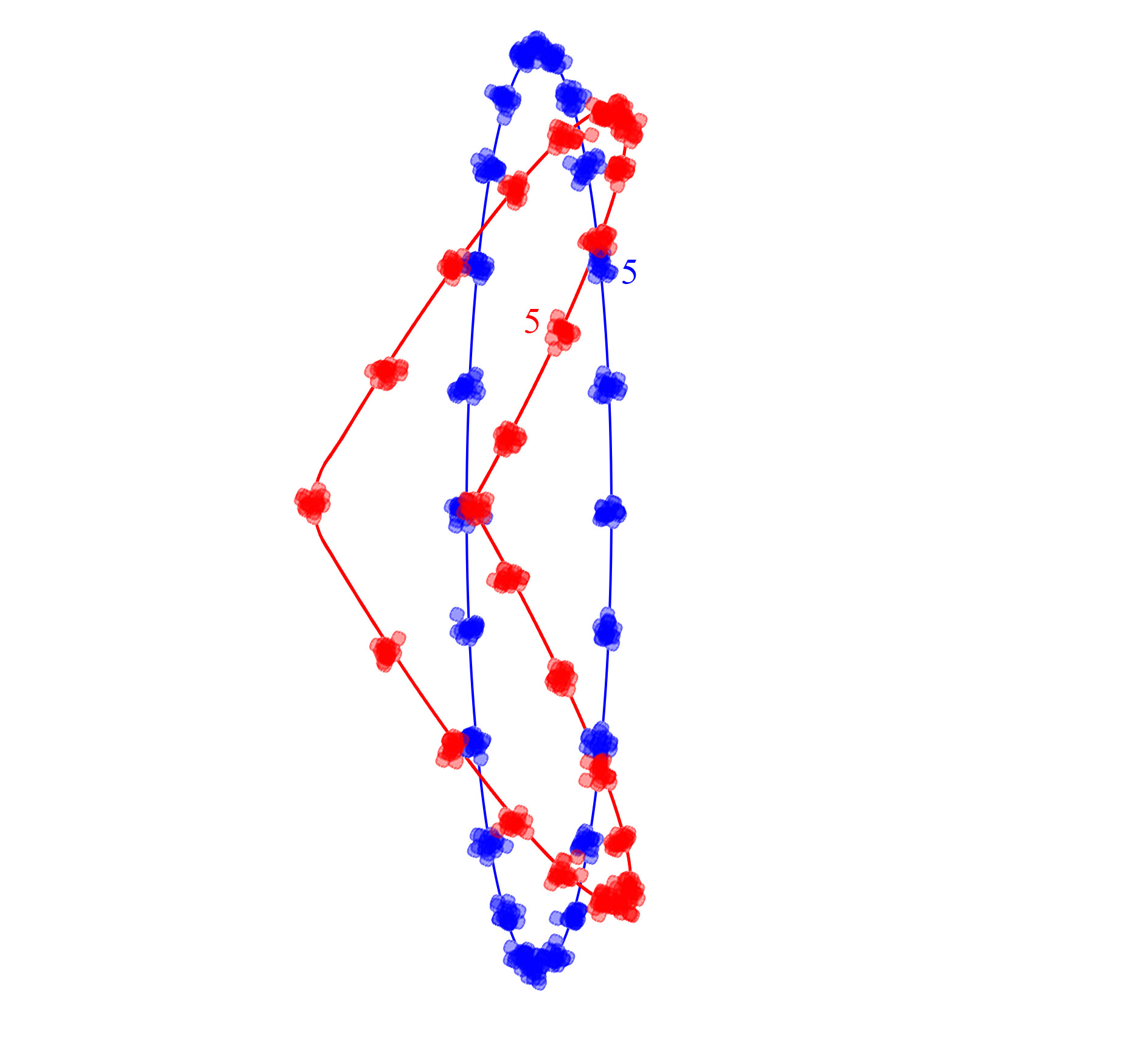}}
         \caption{\centering\footnotesize{GPA alignment}}
         \label{fig:arms_pdm_1}
     \end{subfigure}
     \hfill
     \begin{subfigure}[b]{0.327\textwidth}
         \centering
         \boxed{\includegraphics[width=0.95\textwidth]{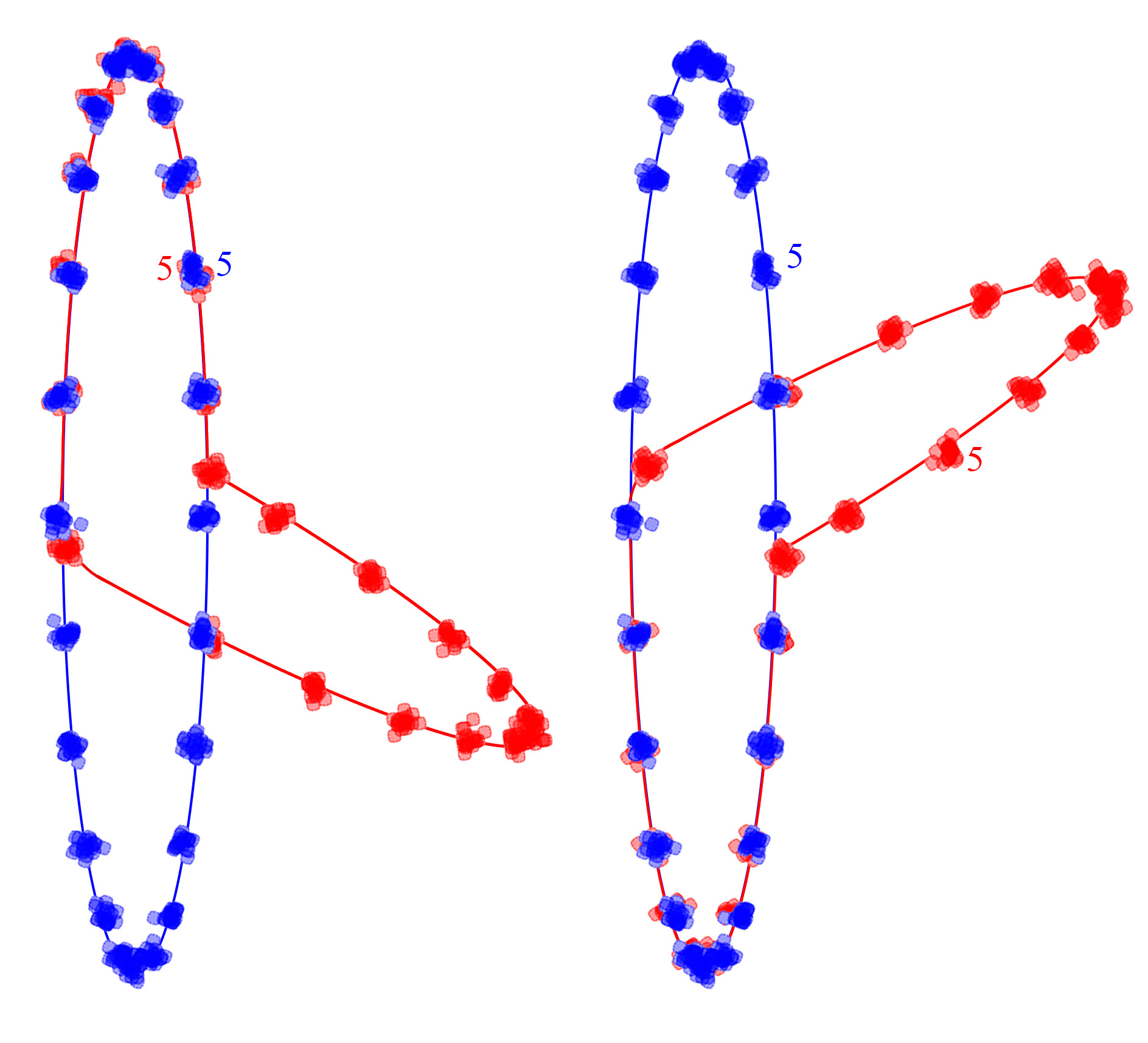}}
         \caption{\centering\footnotesize{Weighted GPA}}
         \label{fig:arms_pdm_2}
     \end{subfigure}
     \caption{\centering\footnotesize{Problem of false positives due to alignment. (a) Red and blue indicate two populations of PDMs. Small crosses are the mean centroids. (b,c) Separation of corresponding local distributions.}}
     \label{fig:arms_pdm}
\end{figure}

 
There are fundamental concerns with the alignment during the preprocessing step. 
\Cref{fig:arms_pdm} illustrates some of the concerns with a simple example. \Cref{fig:arms_pdm_0} shows two ellipsoidal shapes. The blue shape is an ellipsoid, and the red shape is like a boomerang. By visual inspection, we notice that they are different. However, if we think of it as an open arm (blue) and closed arm (red), where each arm consists of three separate parts, namely the upper arm, elbow, and forearm, we observe only a difference around the elbow. The upper and forearm are remaining unchanged. Our visual cortex tries to understand the two objects as a whole, i.e., in a \textit{global coordinate system} (GCS) , even though the red shape in \Cref{fig:arms_pdm_0} is only a locally deformed version of the blue shape. In other words, both shapes are identical at the top and bottom parts but different only at the middle. Now, let us assume both shapes are sampled at 24 positions. By adding independent random noise to each sample point, we simulated 20 PDMs as depicted in \Cref{fig:arms_pdm_0}.
In \Cref{fig:arms_pdm_1}, shapes are aligned without scaling by GPA. The distribution of almost all of the corresponding points (e.g., point No. 5) are remarkably separated that leads to a large number of false positives in mean difference hypothesis testing of local distributions. This phenomenon can also be observed in the analysis of brain structures like the hippocampus, where alignment leads to false positives between CG and treatment groups, further discussed in \Cref{sec:results}. In \Cref{fig:arms_pdm_2}, shapes are aligned based on top and bottom parts with weighted GPA \citep[Sect.~7.6.3]{dryden2016statistical}. Although, it seems weighted GPA is more reasonable than GPA, defining a suitable covariance structure for weighted GPA is not explicit.\par
So far, we have exposed some problems caused by alignment and by understanding shapes in a GCS. We propose a novel hierarchical shape parameterization based on local coordinate systems to overcome the problems. The local parameterization has three major contributions. First, because the representation is translation and rotation invariant, the fundamental issues of alignment are avoided. Second, it understands shapes locally. For example, in \Cref{fig:arms_pdm_0}, we would only detect differences in the middle where deformation occurred. Third, it naturally supports the interpretation of potential shapes differences, e.g., as bending or twisting. Forth, it facilitates skeletal deformation and simulation.\par
Basically, local frames can be defined for different types of object representation. However, a representation that is in particular very suitable is ds-rep. Thus, in this manuscript, we focus on introducing local frames with the application for ds-rep. The paper is structured as follows. Basic notations and amenities of ds-reps are summarized in \Cref{sec:s-rep}. \Cref{fig:ds-rep1} shows a ds-rep of a hippocampus.
\Cref{sec:Conventional_parameterization} introduces the conventional definition of s-rep and ds-rep. The conventional definition is based on GCS with the discussed challenges. In \Cref{sec:local_frames} and \Cref{sec:Local_parameterization}, we introduce the proposed LP-ds-rep. The proposed hierarchical local parameterization of ds-rep, called LP-ds-rep, extracts GOPs with more details and supports sensitive hypothesis testing that is not biased by prior alignment. \Cref{sec:hypothesis_testing} explains euclideanization procedure of spherical data by \textit{principal nested spheres} (PNS) \citep{jung2012analysis}, and adapt the non-parametric hypothesis testing method plus controlling false positive from \citep{Styner2006}. \Cref{sec:simulation} discusses skeletal deformation and simulation by LP-ds-rep. In \Cref{sec:results}, we study hippocampal differences between a group of patients with early Parkinson's disease (PD) and CG given both parameterizations. Besides, we compare the results and show the advantages of our method on simulated data. Finally, we summarize and conclude the work in \Cref{sec:discussion}.
\section{Skeletal representation}\label{sec:s-rep}
The m-rep and its properties have been extensively studied in the literature \citep{pizer1999segmentation,fletcher2004principal,siddiqi2008medial} all the way back to Blum's medial structure \citep{blum1967transformation}. \Cref{fig:m-rep1} illustrates 2D m-reps for the previously discussed shapes with smooth boundaries. Briefly, an m-rep is a sample set of \textit{medial axis} and \textit{spokes}. The medial axis of object $\Omega$ is $M_\Omega=\{\bm{p}\in\Omega\setminus\mathcal{B}\mid\;\mid\{\bm{q}\in\mathcal{B}\mid\,\|\bm{p}-\bm{q}\|=d_{min}(\bm{p},\mathcal{B})\}|_c>1 \}$ where $\mathcal{B}$ is the boundary of $\Omega$, $d_{min}(\bm{p},\mathcal{B})$ is the minimum Euclidean distance between $\bm{p}$ and $\mathcal{B}$, and $|\;|_c$ is the cardinality sign. In other words, the medial axis is centers of all inscribed spheres of $\Omega$, bi-tangent or multi-tangent to $\mathcal{B}$. A spoke is a vector connecting the center of an inscribed sphere to $\mathcal{B}$ where its length is equal to the sphere's radius. Thus, an m-rep reflects the interior object properties such as local widths and directions. However, as pointed out in \citep{pizer2013nested}, the m-rep is sensitive to boundary noise because every protruding boundary kink results in additional medial branches. This sensitivity affects m-rep correspondence among a population as two versions of the same objects can result in significantly different m-reps. Thus, \citep{pizer2013nested} relaxed the mentioned condition and defined s-rep. As described in \citep{liu2021fitting}, for a slabbed-shaped object like the hippocampus, an s-rep has the form $(M,S)$, where $M$ is a smooth ellipsoidal skeletal sheet and $S$ is the field of non-crossing vectors (spokes) on $M$. The tail of each spoke $\bm{s}\in{S}$ is at a point $\bm{p}\in{M}$ and its tip is at the object boundary $\mathcal{B}(\bm{p})$. $(M,S)$ can be assumed as a combination of three components $(M_+,S_+)$, $(M_-,S_-)$, and $(M_0,S_0)$ such that $M_0$ is a fold curve divides $M$ into two collocated submanifolds $M_+$ and $M_-$. $S_+$ and $S_-$ map $M_+$ and $M_-$ respectively to two sides of the object's boundary considered as northern and southern part. Also, $S_0$ maps $M_0$ to the crest part of the boundary. We call a spoke $\bm{s}$ an \textit{up spoke}, \textit{down spoke}, or \textit{crest spoke} if it belongs to $S_+$, $S_-$, or $S_0$ respectively.\par
The relaxed conditions assure stability in the branching structure and thus good case-to-case correspondence across a population of s-reps. In practice, we sample a finite number of corresponding spokes from an s-rep to obtain a ds-rep. The conventional ds-rep parameterization is understood in a GCS explained in more detail in \Cref{sec:Conventional_parameterization}. Afterward, a novel parameterization based on a hierarchical structure of the local frames is introduced. Also, we name the conventional parameterization as \textit{globally parameterized ds-rep} (GP-ds-rep), and the new parameterization as \textit{locally parameterized ds-rep} (LP-ds-rep). Further, $s^{\GP}$ and $s^{\LP}$ denote GP-ds-rep and LP-ds-rep respectively.

\subsection{Parameterizations}\label{sec:ds-rep_parameterization}
\subsubsection{GP-ds-rep}\label{sec:Conventional_parameterization}
There are different ways to fit and parameterize a ds-rep. A current implementation described in detail by \citep{liu2021fitting} is available under the open-source toolbox SlicerSALT (\url{http://salt.slicer.org}). A GP-ds-rep is a tuple $s^{\GP}=(\bm{p}_i,\bm{u}_i,r_i)$ where $\bm{p}_i\in\mathbb{R}^3$, $\bm{u}_i\in{\mathbb{S}^2}$ and $r_i\in\mathbb{R}^+$ are $i$th spoke's tail position, direction, and length respectively where $\mathbb{S}^d=\{\bm{x}\in\mathbb{R}^{d+1}|\,\|\bm{x}\|=1\}$ is the unit sphere in arbitrary dimension $d\geq3$, $i=1,...,n_s$, and $n_s$ is the number of spokes. Based on the current model fitting, some spokes share a common tail position, so we have $s^{\GP}=(\bm{p}_j,\bm{u}_i,r_i)$ where $j=1,...,n_p$, and $n_p<n_s$. The set $\{\bm{p}_j\}_{j=1}^{n_p}$ forms an $n_p\times{3}$ configuration matrix $P$ representing the skeletal PDM. Assume $I_{n_p}$ as ${n_p}\times{n_p}$ identity matrix and $\bm{1}_{n_p}$ as ${n_p}\times{1}$ vector of ones. Location and scale can be removed by centering and normalizing skeletal PDM to obtain pre-shape $\tilde{P}=\frac{C_{n_p}P}{\|C_{n_p}P\|}$, where $C_{n_p}=I_{n_p}-\frac{1}{n_p}\bm{1}_{n_p}{\bm{1}_{n_p}^T}$ is the centering matrix, and $\|X\|=\sqrt{trace(X^TX)}$ is the Euclidean norm. Since $\|\tilde{P}\|=1$, the pre-shape $\tilde{P}$ lives on the hypersphere $\mathbb{S}^{3n_p-1}$ \citetext{\citealp[Ch.2]{dryden2016statistical}; \citealp{schulz2016non}}.
Thus a GP-ds-rep lives on a manifold as a direct product of Riemannian symmetric spaces, namely $s^{\GP} \in \mathbb{S}^{3n_p-1}\times(\mathbb{S}^2)^{n_s}\times\mathbb{R}_{+}^{n_s+1}$ where $\mathbb{S}^{3n_p-1}$ indicates the pre-shape space of the skeletal PDM, $(\mathbb{S}^2)^{n_s}$ is the space of $n_s$ spokes' directions, and $\mathbb{R}_{+}^{n_s+1}$ is the space of spokes' lengths and the scaling factor. Note that spoke positions and directions in this parameterization are in a GCS independent of the skeletal sheet structure. This lead to the discussed challenges in statistical analysis because alignment is necessary.\par
For m-rep, a semi-local parameterization was proposed by \citep{fletcher2003statistics} based on local frames $(\bm{n},\bm{b},\bm{b}^{\perp})\in{SO(3)}$, where $\bm{n}$ is normal to the medial sheet $M$ at $\bm{p}\in{M}$, $\bm{b}=\frac{\bm{u}_1+\bm{u}_2}{\|\bm{u}_1+\bm{u}_2\|}$ is the bisector direction of two equal-length spokes with common position, $\bm{b}^\perp=\bm{n}\times\bm{b}$, and $SO(3)$ is the 3D rotation group. Spokes' directions are defined relative to the local frames by the angle $\theta\in[0,\pi)$ between $\bm{b}$ and the spokes (see \Cref{fig:local_frame_GP}).\par
Because the direction of $\bm{b}$ and $\bm{b}^{\perp}$ depends on the spokes' directions, if $\theta=\frac{\pi}{2}$ then $\bm{b}$ takes an arbitrarily direction that violates the uniqueness and consistency of the fitted frame. Besides, the spokes' tail positions and frame directions are in GCS. 
Thus for statistical shape analysis, pre-alignment is still necessary.

Inspired by both, Cartan's moving frames on space curves \citep{alma991001580569705596}, and Fletcher's semi-local parametrization, we propose a fully local ds-rep parameterization. By utilizing the inherent hierarchical structure of ds-reps, we provide a consistent definition of local frames that is independent of GCS and also avoids arbitrarily frame rotation. This can be done by introducing a leaf-shaped skeletal structure of the skeletal sheet according to the ellipsoidal design of the ds-rep applied in \citep{liu2021fitting} (see \Cref{fig:Leaf-Shape_And_LP-ds-rep}(Top)). 

\subsubsection{Local frames}\label{sec:local_frames}
For GP-ds-rep fitting, \citep{liu2021fitting} first deformed an object to an ellipsoid by mean curvature flow. Then by inverse mean curvature flow, warped the ellipsoid's GP-ds-rep to the target object. Finally, spokes are refined such that the implied boundary becomes as close as possible to the real boundary. As a result, the implied boundary corresponds to the ellipsoid's boundary, and the skeletal sheet corresponds to the ellipsoid's skeletal sheet. In other words, the structure of the fitted GP-ds-rep is associated with the ellipsoid's GP-ds-rep i.e., ellipsoid's medial axis. By assuming a good correspondence between the ellipsoid's medial axis and the object's skeletal sheet, we design a hierarchical structure for the ellipsoid's medial axis and expand it to the object's skeletal sheet. Then on the basis of the obtained structure, we define consistent fitted frames in a population of GP-ds-reps.\par
In this article, we only consider slabular shapes corresponding to an eccentric ellipsoid with principal radii $a,b,c\in\mathbb{R}^+$ such that $a>b>c$. As discussed above, up and down spokes correspond to the ellipsoid’s northern and southern side, while crest spokes correspond to the ellipsoid’s crest.\par
\begin{figure}[ht]
\centering
  \boxed{\includegraphics[width=0.98\textwidth]{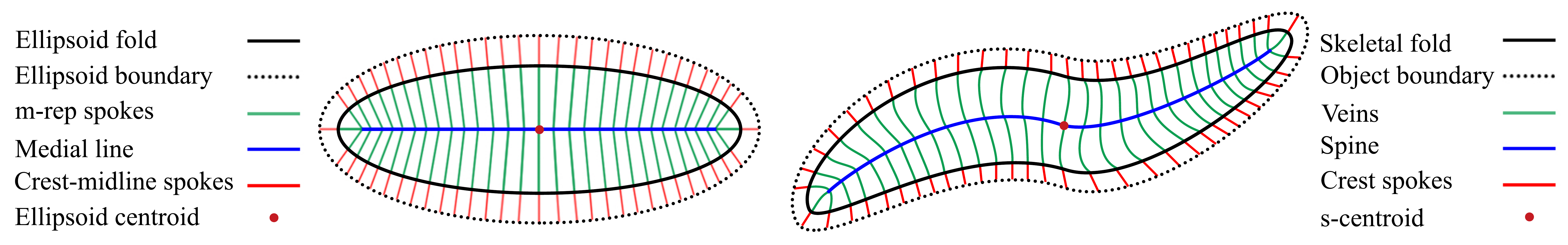}}
\caption{\centering\footnotesize{Ellipsoid's Medial axis deformation. Left: Medial axis of an eccentric 3D ellipsoid. Right: Object's s-rep skeletal sheet.}}
\label{fig:Medial_axis_deformation}
\end{figure}

As illustrated in \Cref{fig:Medial_axis_deformation}(left), the medial axis of an ellipsoid is an ellipse (i.e., a 2D ellipsoid). The ellipse is symmetric and has a symmetric 2D m-rep. The m-rep consists of medial points located on a straight medial line and a set of spokes. The middle point of the medial line is the ellipsoid's centroid (i.e., center of gravity $\bar{\bm{x}}=\frac{1}{n}\sum_{i=1}^n\bm{x}_i$). The boundary of the ellipse is the skeletal sheet's fold (i.e., skeletal edge) of the ellipsoid. The extension of m-rep spokes connecting the fold to the ellipsoid boundary is a subset of crest spokes. We call the members of this subset \textit{crest-midline spokes}. For the model fitting, \citep{liu2021fitting} deformed the ellipsoid to the object. After the deformation, the flat ellipsoid's medial axis transforms to a nonlinear surface as a 2D manifold $M$ (i.e., topologically a 2-dimensional disk). Consequently, straight lines on it (e.g., medial line and m-rep spokes) become curves. Also \citep{liu2021fitting} assumed more or less a diffeomorphic transformation. Thus the generated curves do not cross each other.\par
We call the deformed medial line the \textit{spine}, and deformed m-rep spokes \textit{veins}. Thus, veins are a set of non-crossing curves emanating from the spine. During the deformation, the ellipsoid's centroid moves and ultimately rests in the middle of the object. We assume the displaced centroid as an intrinsic centroid, and call it \textit{skeletal centroid} or \textit{s-centroid}.
\Cref{fig:Medial_axis_deformation} provides an intuition about the ellipsoid's medial axis deformation.\par
\begin{figure}[ht]
     \centering
     \begin{subfigure}[b]{0.47\textwidth}
         \centering
         \boxed{\includegraphics[width=0.97\textwidth]{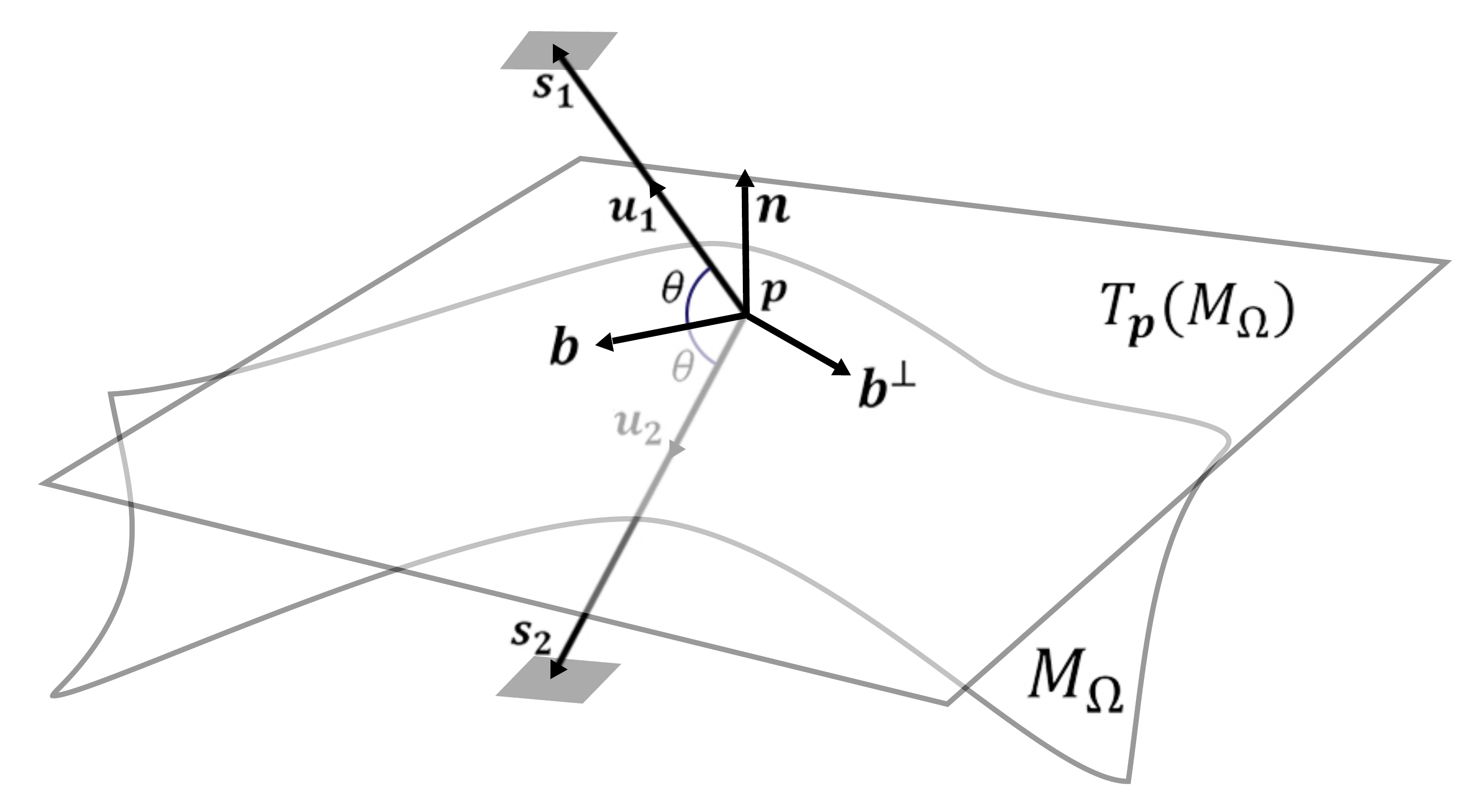}}
         \caption{\centering\footnotesize{m-rep frame}}
         \label{fig:local_frame_GP}
     \end{subfigure}
     \hfill
     \begin{subfigure}[b]{0.47\textwidth}
         \centering
         \boxed{\includegraphics[width=0.97\textwidth]{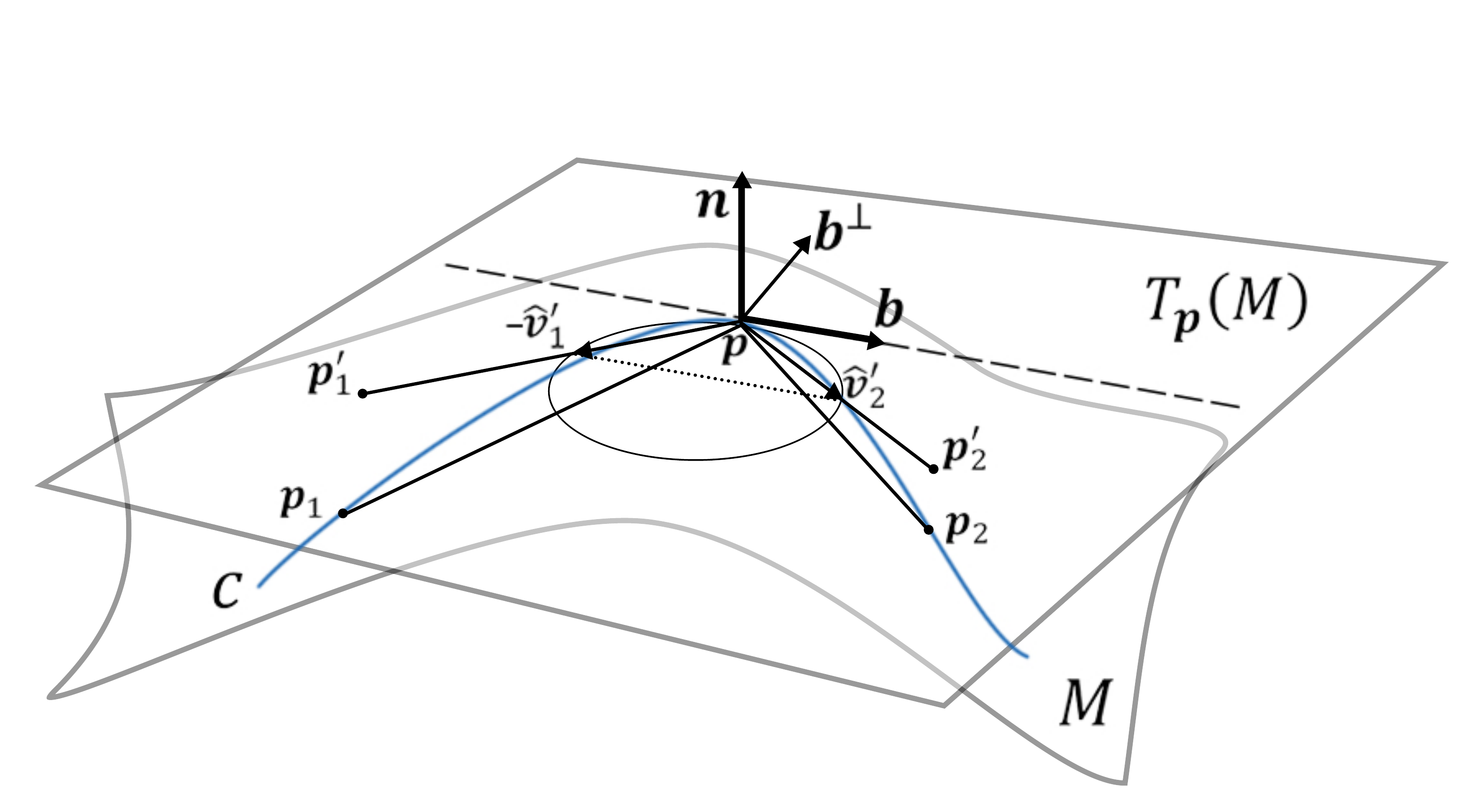}}
         \caption{\centering\footnotesize{LP-ds-rep frame}}
         \label{fig:local_frame_LP}
     \end{subfigure}
     \caption{\centering\footnotesize{Illustration of a local frames. $\bm{n}$ is normal of tangent planes $T_{\bm{p}}(M)$ and $T_{\bm{p}}(M_\Omega)$. (a) $\bm{s}_1$ and $\bm{s}_2$ are equal-length spokes with unit directions $\bm{u}_1$ and $\bm{u}_2$, and $\bm{b}=\frac{\bm{u}_1+\bm{u}_2}{\|\bm{u}_1+\bm{u}_2\|}$ (b) $c$ is a smooth curve on $M$. $-\bm{p}'_1$ and $\bm{p}'_2$ are the projection of $\bm{p}_1$ and $\bm{p}_2$ on $T_{\bm{p}}(M)$. $\hat{\bm{v}}'_1=\frac{\bm{p}-\bm{p}'_1}{\|\bm{p}-\bm{p}'_1\|}$, $\hat{\bm{v}}'_2=\frac{\bm{p}'_2-\bm{p}}{\|\bm{p}'_2-\bm{p}\|}$, and $\bm{b}=\frac{\hat{\bm{v}}'_2+\hat{\bm{v}}'_1}{\|\hat{\bm{v}}'_2+\hat{\bm{v}}'_1\|}$.}}
\label{fig:local_frame}
\end{figure}

Let $c\in{M}$ be a smooth curve in $\mathbb{R}^3$. We consider $\bm{b}\in{T_{\bm{p}}(M)}$ as the unit velocity vector tangent to $c$ where ${T_{\bm{p}}(M)}$ is the local tangent plane of $M$ at $\bm{p}\in{c}$ with normal $\bm{n}$. The local frame can be defined as $(\bm{n},\bm{b},\bm{b}^\perp)\in{SO(3)}$ where $\bm{b}^\perp=\bm{n}\times\bm{b}$ (see \Cref{fig:local_frame_LP}).\par
Note that in practice, the directions of $\bm{b}$ and $\bm{n}$ may choose two opposite directions because $M$ is double-sided. So far, the frame directions are in GCS. To have a consistent frame definition independent of GCS, we design a hierarchical structure.\par
We start with the m-rep of the ellipsoid's medial axis and prioritize the ellipsoid's centroid, medial line, and spokes, respectively. Also, we prioritize points on the medial line closer to the ellipsoid's centroid and points on the spokes closer to the medial line. Analogous to the m-rep, we give priority to the s-centroid, spine, and veins, respectively. Further, based on the geodesic distance on curves, we prioritize spinal points closer to the s-centroid and points on veins closer to the spine. Thus, given a frame at each skeletal point, we introduce a hierarchical frame structure.\par
Except for the s-centroid frame, each frame has a prior frame called \textit{parent frame}. In GP-ds-rep, we have a finite number of prioritized frames on the spine or a vein. A vector that connects a frame to its parent frame is called \textit{connection}. The tip of a connection is at the frame's origin, and its tail is at the parent's origin. Therefore, like a spanning tree, each frame has a parent but may have multiple children. Further, we assume that the s-centroid frame is the parent of itself with the connection $\bm{0}=(0,0,0)^T$.\par
We approximate the direction of $\bm{b}$ at point $\bm{p}\in{M}$ based on three consecutive frames. Except for the s-centroid frame and two critical endpoints of the spine that we will explain later, each spinal frame has a spinal parent frame and a spinal child frame. Let $\bm{p}_1$ and $\bm{p}_2$ be the position of the parent and the child frame of $\bm{p}$. As illustrated in \Cref{fig:local_frame}, assume $\bm{v}_1=\bm{p}-\bm{p}_1$ and $\bm{v}_2=\bm{p}_2-\bm{p}$ as connections. Let $\bm{p}'_1$ and $\bm{p}'_2$ be the projection of $\bm{p}_1$ and $\bm{p}_2$ on $T_{\bm{p}}(M)$, respectively. We consider $\bm{b}=\frac{\hat{\bm{v}}'_2+\hat{\bm{v}}'_1}{\|\hat{\bm{v}}'_2+\hat{\bm{v}}'_1\|}$ where $\hat{\bm{v}}'_1=\frac{\bm{p}-\bm{p}'_1}{\|\bm{p}-\bm{p}'_1\|}$, and $\hat{\bm{v}}'_2=\frac{\bm{p}'_2-\bm{p}}{\|\bm{p}'_2-\bm{p}\|}$. In this sense, $\bm{b}$ is a unit vector tangent to a circle (or a line) crossing $\bm{p}-\hat{\bm{v}}'_1$, $\bm{p}$, and $\bm{p}+\hat{\bm{v}}'_2$.\par
\begin{figure}[ht]
\centering
  \boxed{\includegraphics[width=0.9\textwidth]{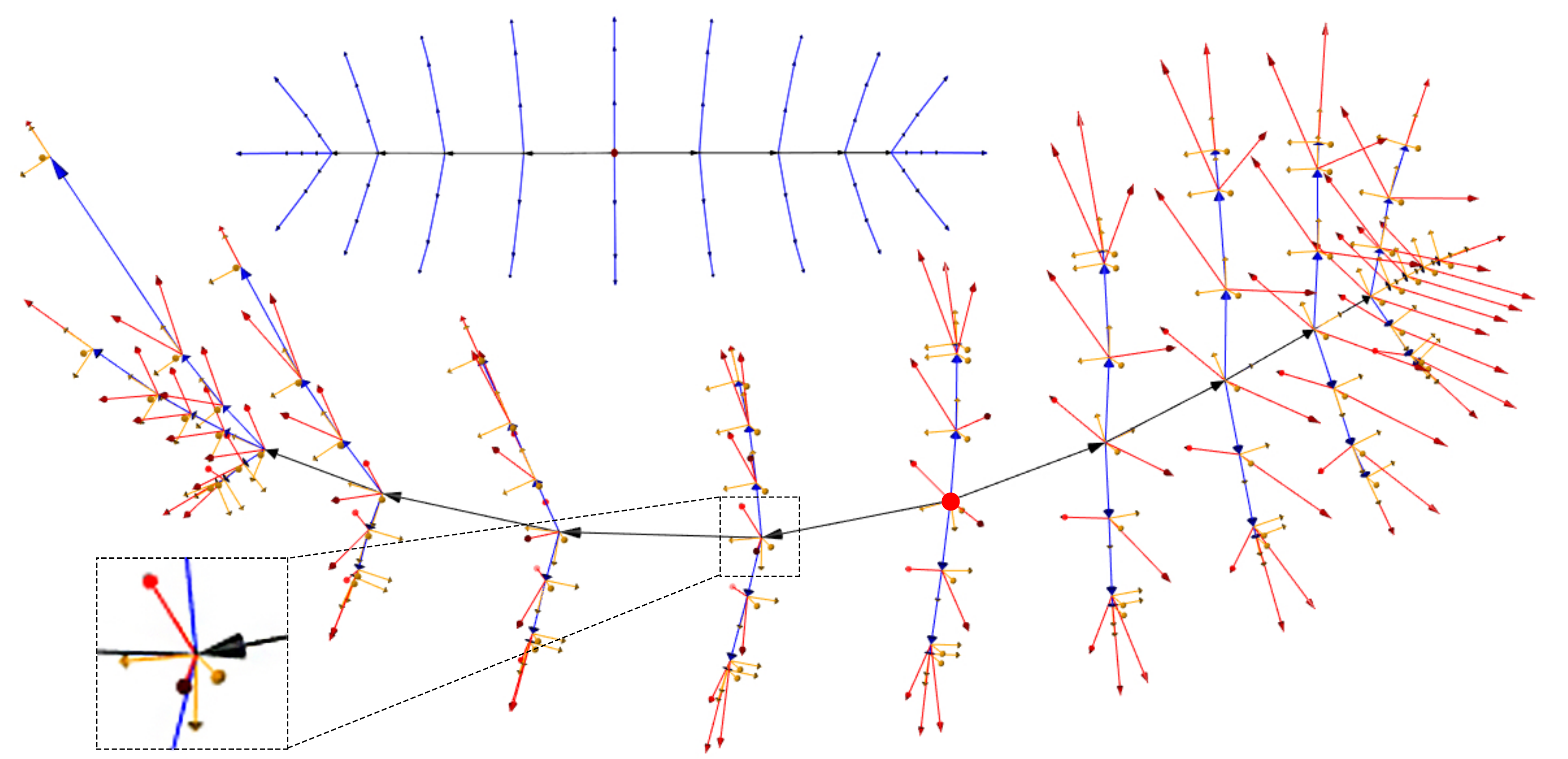}}
\caption{\centering\footnotesize{LP-ds-rep. Top: Hierarchical structure of the ellipsoid's medial axis. Black and blue arrows are connections on the medial line and m-rep spokes. Red dot is the ellipsoid's centroid. Bottom: A fitted LP-ds-rep to a hippocampus. Red arrows indicate spokes. Black and blue arrows are connections on the spine and veins. Orange arrows depict orthogonal local frames. Red dot is the s-centroid.}}
\label{fig:Leaf-Shape_And_LP-ds-rep}
\end{figure}

The endpoints of the spine are critical because their frames have no children on the spine. By construction, the m-rep medial line is a part of the ellipse's major axis. After deformation, the major axis becomes a curve we call \textit{major curve}. The major curve contains the spine and two veins. We consider the closest skeletal point (in geodesic sense) on these veins to the spine as the spine's extension and treat the critical points as any other spinal point. The s-centroid frame has two spinal children. Let $\bm{p}_1$ and $\bm{p}_2$ be the position of the children. We define $\bm{b}=\frac{\hat{\bm{v}}'_2-\hat{\bm{v}}'_1}{\|\hat{\bm{v}}'_2-\hat{\bm{v}}'_1\|}$, where $\hat{\bm{v}}'_1=\frac{\bm{p}-\bm{p}'_1}{\|\bm{p}-\bm{p}'_1\|}$, and $\hat{\bm{v}}'_2=\frac{\bm{p}'_2-\bm{p}}{\|\bm{p}'_2-\bm{p}\|}$. Since a vein frame has a parent and a child on the same vein, we consider the same definition for them as discussed for spinal frames. Note we treat a vein frame at the intersection of a vein and the spine as a spinal frame. For the frames on crest spokes' tails (i.e., on the skeletal fold), we assume the tip of the crest spokes as the position of the child frames. The same procedure is applicable for the ellipsoid's GP-ds-rep.
\Cref{fig:Leaf-Shape_And_LP-ds-rep} illustrates the hierarchical structure and a fitted LP-ds-rep to a left hippocampus as described in the next section.\par
\subsubsection{LP-ds-rep}\label{sec:Local_parameterization}
Given the fitted hierarchical frame structure introduced in the previous section, we are now in the position to define LP-ds-rep. In an LP-ds-rep, spokes and connections are measured based on their local frames, i.e., their tails are located at the origin of a frame.
Let $\bm{u}_i$ and $\bm{v}_j$ be the $i$th spoke direction and $j$th connection direction in GCS respectively, where $i=1,...,n_s$ and $j=1,...,n_p$. Consequently, we denote $\bm{u}^\ast_{i}$ and $\bm{v}^\ast_{j}$ as spoke and connection directions based on their local frame, i.e. we re-parameterize $\bm{u}_i$ and $\bm{v}_j$ to $\bm{u}^\ast_{i}$ and $\bm{v}^\ast_{j}$ respectively. Similarly, if $F_j=(\bm{n}_j,\bm{b}_j,\bm{b}_j^\perp)$ be the frame $F_j$ in GCS then $F_j^{\ast}=(\bm{n}^{\ast}_j,\bm{b}^{\ast}_j,\bm{b}^{\ast\perp}_j)$ denotes $F_j$'s vectors, based on its parent frame.\par
To calculate a vector direction according to a local frame, we use the spherical rotation matrix $R(\bm{x},\bm{y})=I_3+(\sin{\alpha})(\bm{y}\bm{w}^T-\bm{w}\bm{y}^T)+(\cos{\alpha}-1)(\bm{y}\bm{y}^T+\bm{w}\bm{w}^T)$, where $\bm{x},\bm{y}\in{\mathbb{S}^2}$, $\bm{w}=\frac{\bm{x}-\bm{y}(\bm{y}^T\bm{x})}{\|\bm{x-\bm{y}(\bm{y}^T\bm{x})}\|}$ and $\alpha=\cos^{-1}(\bm{y}^T\bm{x})$. Therefore, $R(\bm{x},\bm{y})$ transfers $\bm{x}$ to $\bm{y}$ along the shortest geodesic and we have $R(\bm{x},\bm{y})\bm{x}=\bm{y}$ \citep{amaral2007pivotal}.\par
For example, let frame $F^\dagger=(\bm{n},\bm{b},\bm{b}^{\perp})$ be the parent of $\tilde{F}$, both in GCS. Let $\bm{e}_1=(1,0,0)^T$, $\bm{e}_2=(0,1,0)^T$, and $\bm{e}_3=(0,0,1)^T$ be the axes unit vectors of GCS. We align $F^\dagger$ to $\tilde{I}=(\bm{e}_3,\bm{e}_1,\bm{e}_2)$ such that $R_2R_1F^\dagger=\tilde{I}$, where $R_1=R(\bm{n},\bm{e}_3)$, and $R_2=R(R_1\bm{b},\bm{e}_1)$. Thus, $\tilde{F}^{\ast}=R_2R_1\tilde{F}$ represents $\tilde{F}$ in its parent coordinate system. In case we obtain $R_2R_1F^\dagger=(\bm{e}_3,\bm{e}_1,-\bm{e}_2)$, we adjust the result by $R_2R_1\tilde{F}(\bm{1}_3,\bm{1}_3,-\bm{1}_3)$ because $R_2R_1F^\dagger(\bm{1}_3,\bm{1}_3,-\bm{1}_3)=\tilde{I}$ where $\bm{1}_3=(1,1,1)^T$. Note that frame vectors are orthogonal, so after rotating $\bm{n}$ to the north pole by $R_1$, the shortest geodesic between $\bm{b}$ and $\bm{e}_1$ would be on the equator. This preserve the direction of $R_1\bm{n}$ while $R_2$ rotates $R_1\tilde{F}$.\par
We follow the the same procedure to calculate the spokes' and connections' directions based on their local frames $F^{\ast}_j$. Finally, a LP-ds-rep is given by $s^{\LP}=(\bm{u}^{\ast}_i,r_i,F^{\ast}_j,\bm{v}^{\ast}_j,v_j)$, where $\bm{u}^{\ast}_i\in{\mathbb{S}^2}$ and $\bm{v}^{\ast}_j\in{S}^2$ are $i$th and $j$th spoke direction and connection direction relative to their local frame with lengths $r_i\in{\mathbb{R}}^+$ and $v_j\in{\mathbb{R}^+}$ respectively, $F^{\ast}_j=(\bm{n}_j^{\ast},\bm{b}_j^{\ast},\bm{b}_j^{\ast\perp})\in{SO(3)}$ is the $j$th frame in its parent coordinate system, and $i=1,...,n_s$ and $j=1,...,n_p$ where $n_s$ is the number of spokes and $n_p$ is the number of frames.\par
Thus, by construction, the LP-ds-rep is invariant under the act of rigid similarity transformation (i.e., rotation and translation). To remove the scale, we define \textit{LP-size} as $\ell=\sum_{i=1}^{n_s}r_i+\sum_{j=1}^{n_p}v_j$. A scaled LP-ds-rep can be expressed by $s^{\LP}=(\bm{u}^{\ast}_i,\rho_i,F^{\ast}_j,\bm{v}^{\ast}_j,\tau_j)$, where $\rho_i=\frac{r_i}{\ell}$, and $\tau_j=\frac{v_j}{\ell}$. Thus, the LP-size of an scaled LP-ds-re is equal to one. Recall, for a GP-ds-rep, the \textit{GP-size} is defined as the centroid size e.g., centroid size of spokes' tails. Note the centroid is an extrinsic property, and the centroid size might be a poor measure for the size of an object. Intuitively, by opening or closing an arm, the arm's volume remains the same despite its centroid size, i.e., the closed arm has a smaller centroid size as its boundary points are closer to the centroid (see \Cref{fig:arms_pdm_0}).\par
As \Cref{sec:Conventional_parameterization} discussed, the GP-ds-rep space is $\mathbf{S}^{\GP}=\mathbb{S}^{3n_s-1}\times(\mathbb{S}^2)^{n_s}\times\mathbb{R}_{+}^{n_s+1}$. In LP-ds-rep, we do not have any pre-shape space. LP-ds-rep GOPs are directions and lengths of the vectors (i.e., spokes, connections, and frames) plus LP-size. Thus the space is $\mathbf{S}^{\LP}=(\mathbb{S}^2)^{n_s+4n_p}\times\mathbb{R}_{+}^{n_s+n_p+1}$, where $(\mathbb{S}^2)^{n_s+4n_p}$ is the space of directions and $\mathbb{R}_{+}^{n_s+n_p+1}$ is the space of vectors' lengths and LP-size. If we euclideanize directions as described in the following sections, the space of euclideanized LP-ds-rep is $(\mathbb{R}^2)^{n_s+4n_p}\times\mathbb{R}_{+}^{n_s+n_p+1}$.
\subsection{Population mean}\label{sec:mean}
Having a population of ds-reps, suitable methods to calculate means are required in order to perform hypothesis tests on mean differences. The corresponding method should incorporate all geometrical components of the model. Both shape spaces, the GP-ds-rep space, and the LP-ds-rep space are composed of several spheres and a real space. This section will first discuss an approach to analyze the spherical parts by PNS. Afterward, approaches to produce GP-ds-rep means and LP-ds-rep means are discussed.
\subsubsection{PNS}\label{sec:PNS}
PNS \citep{jung2012analysis} estimates the joint probability distribution of data on a d-dimensional sphere $\mathbb{S}^d$ by a backward view, i.e., in decreasing dimension. Starting with $\mathbb{S}^d$, PNS fits the best lower-dimensional subsphere in each dimension. A subsphere is called \textit{great subsphere} if its radius is equal to one; otherwise, it is called \textit{small subsphere}. For the special case $d=2$, it is called \textit{great circle} or \textit{small circle} respectively. To choose between the great or small subsphere, we use the Kurtosis test from \citep{kim2020kurtosis}.\par
\begin{figure}[ht]
   \begin{minipage}{0.48\textwidth}
     \centering
     \boxed{\includegraphics[width=.97\linewidth]{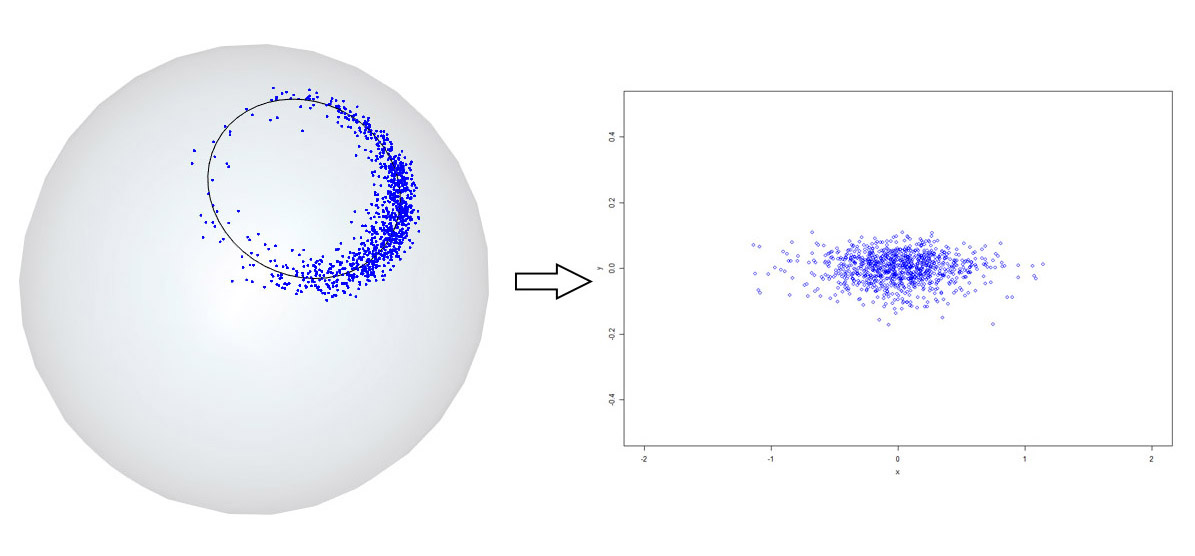}}
     \caption{\centering\footnotesize{ PNS euclideanization. Left: Small circle distribution and the fitted circle on $\mathbb{S}^2$. Right: Euclideanizated data.}}
     \label{fig:PNS_Euclideanization}
   \end{minipage}
   \begin{minipage}{0.48\textwidth}
     \centering
     \boxed{\includegraphics[width=.97\linewidth]{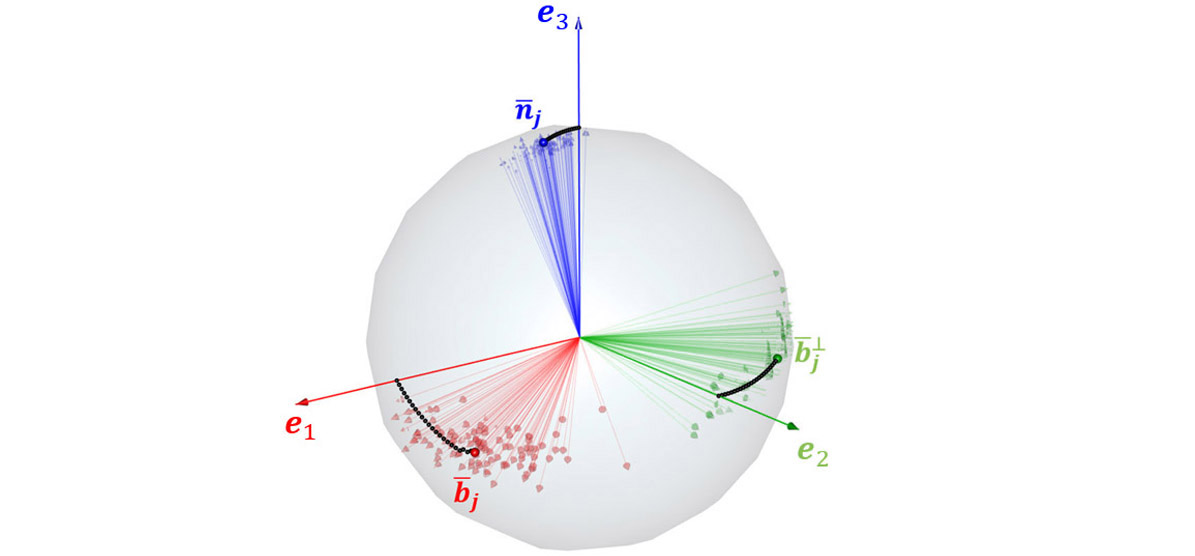}}
     \caption{\centering\footnotesize{Mean frame by gradient descent with initial frame $\tilde{I}$. Black dots show the movement of $\tilde{I}$ toward the Fr\'echet means of frames' vectors.}}
     \label{fig:mean_frame2}
   \end{minipage}\hfill
\end{figure}

PNS is designed for spherical distributions and in particular for small sphere distributions as described in \citep{kim2019small}. PNS captures the curviness of circular distributions and euclideanize data as residuals. The PNS residuals on $\mathbb{S}^2$ consist of the geodesic distances between the observations and the fitted circle and the minimal arc length between projected data on the fitted circle to the PNS mean. Therefore in many cases, the distribution of euclideanized data (i.e., residuals) is similar to the bivariate normal distribution. \Cref{fig:PNS_Euclideanization} illustrates fitted circle to a cluster of 1000 observations on $\mathbb{S}^2$, and the PNS residuals. Random points are generated from small sphere distribution $X\sim{f_{S2}(\mu_0,\mu_1,\kappa_0,\kappa_1)}$ \citep{kim2019small} where $\mu_0=(0,0,1)^T$, $\mu_1=(\cos{\frac{\pi}{3}},0,\sin{\frac{\pi}{3}})^T$, $\kappa_0=500$, and $\kappa_1=2$.\par
Alternatively, a simpler but faster euclideanization is to map the data on the tangent space. We transform observations to the north pole $\bm{q}=(0,0,1)^T$ by $R(\bm{\mu}_F,\bm{q})$, where $\bm{\mu}_F$ is the Fr\'echet mean. Then, we map the transformed data to the tangent space $T_{\bm{q}}(\mathbb{S}^2)$ by the Log map $\textup{Log}_{\bm{q}}(\bm{v})=\left(v_1.\frac{\theta}{\sin{\theta}},v_2.\frac{\theta}{\sin{\theta}}\right)^T$, where $\bm{v}=(v_1,v_2,v_3)^T\in\mathbb{S}^2$, and $\theta=\cos^{-1}(\bm{v}^T\bm{q})$ \citep{fletcher2004principal}. Note that for concentrated von Mises-Fisher distribution, the distribution of projected data to the tangent space is close to the distribution of PNS residuals.
\subsubsection{Mean GP-ds-rep}\label{sec:mean_GP-ds-rep}
A method to produce means and shape distributions of a population of GP-ds-reps is \textit{composite PNS} (CPNS) introduced by \citep{pizer2013nested}. The method consists of two steps. First, the two spherical parts of the GP-ds-rep shape space $\mathbb{S}^{3n_s-1}\times(\mathbb{S}^2)^{n_s}\times\mathbb{R}_{+}^{n_s+1}$ are analyzed by PNS. Spokes' lengths and scaling factor can be mapped to $\mathbb{R}^{n_s+1}$ with the $log$. Afterward, all euclideanized variables are concatenated in addition to some scaling factors that make all variables commensurate. The covariance structure of the resulting matrix is investigated with PCA. Consequently, the mean GP-ds-rep is defined as the origin of the CPNS space. This method depends on a proper pre-alignment and is computationally expensive because PNS has to fit sequential high dimensional sub-spheres to $\mathbb{S}^{3n_s-1}$.
\subsubsection{Mean LP-ds-rep}\label{sec:mean_LP-ds-rep}
To formalize the estimation of LP-ds-rep mean, first we define LP-ds-rep distance. Let $d_g$ and $d$ be the geodesic and Euclidean distance, respectively. The distance between two scaled LP-ds-reps $s^{\LP}_1=(\bm{u}^{\ast}_{1i},\rho_{1i},F^{\ast}_{1j},\bm{v}^{\ast}_{1j},\tau_{1j})$, and $s^{\LP}_2=(\bm{u}^{\ast}_{2i},\rho_{2i},F^{\ast}_{2j},\bm{v}^{\ast}_{2j},\tau_{2j})$ is given by
\begin{multline}\label{equ:ds-rep_distance}
d_s(s_{1}^{\LP},s_{2}^{\LP})=\left(\sum_{i=1}^{n_s}d_g^2(\bm{u}^{\ast}_{1i},\bm{u}^{\ast}_{2i})+\sum_{i=1}^{n_s}d^2(\rho_{1i},\rho_{2i})+\sum_{j=1}^{n_p}d_F^2(F^{\ast}_{1j},F^{\ast}_{2j})+\right.\\
\left.\sum_{j=1}^{n_p}d_g^2(\bm{v}^{\ast}_{1j},\bm{v}^{\ast}_{2j})+\sum_{j=1}^{n_p}d^2(\tau_{1j},\tau_{2j})\right)^\frac{1}{2},
\end{multline}
where $\forall{j}$, $d_F(F^{\ast}_{1j},F^{\ast}_{2j})=\left(d_g^2(\bm{n}^{\ast}_{1j},\bm{n}^{\ast}_{2j})+d_g^2(\bm{b}^{\ast}_{1j},\bm{b}^{\ast}_{2j})+d_g^2(\bm{b}^{\ast\perp}_{1j},\bm{b}^{\ast\perp}_{2j})\right)^\frac{1}{2}$.\par
If $s_{{1}}^{\LP},...,s_{{N}}^{\LP}$ be a population of scaled LP-ds-reps then mean LP-ds-rep is
\begin{equation}\label{equ:mean_ds-rep}
        \bar{s}^{\LP}=\argmin_{s^{\LP}\in\mathbf{S}^{\LP}}\sum_{k=1}^{N}d_s^2(s^{\LP},s^{\LP}_{{k}}).
\end{equation}
Note in \cref{equ:ds-rep_distance}, $d_g(.)\in[0,\pi]$, and $\forall{i,j}$, $d(\rho_{1i},\rho_{2i})\in[0,1]$, $d(\tau_{1i},\tau_{1i})\in[0,1]$. To commensurate $d(.)$ with $d_g(.)$ we can use linear shift function $f(d(.))=\pi{d(.)}$ instead of $d(.)$. However, this change does not influence the result of \cref{equ:mean_ds-rep} because right-hand-side terms of \cref{equ:ds-rep_distance} are positive and independent. Assume $\bar{s}^{\LP}=(\bar{\bm{u}}^{\ast}_i,\bar{\rho}_i,\bar{F}^{\ast}_j,\bar{\bm{v}}^{\ast}_j,\bar{\tau}_j)$ and $\forall{i,j}$ let
\begin{align}
    \bar{\bm{u}}^{\ast}_i=\argmin_{\bm{u}\in{\mathbb{S}^2}}\sum_{k=1}^{N}d_g^2(\bm{u}^{\ast},\bm{u}^{\ast}_{ik}), \quad &  \bar{\rho}_i=\argmin_{\rho\in{\mathbb{R}^+}}\sum_{k=1}^{N}d^2(\rho,\rho_{ik}), &
    \bar{F}_j=\argmin_{F\in{SO(3)}}\sum_{k=1}^{N}d_F^2(F,F_{jk}), \nonumber \\
    \bar{\bm{v}}^{\ast}_j=\argmin_{\bm{v}\in{\mathbb{S}^2}}\sum_{k=1}^{N}d_g^2(\bm{v}^{\ast},\bm{v}^{\ast}_{jk}), \quad & \bar{\tau}_j=\argmin_{\tau\in{\mathbb{R}^+}}\sum_{k=1}^{N}d^2(\tau,\tau_{jk}) \label{eq:five_optimization}.
\end{align}
By assuming the existence of unique solutions for optimization problems \eqref{eq:five_optimization}, $\bar{\bm{u}}^{\ast}_i$ and $\bar{\bm{v}}^{\ast}_j$ can be estimated as the Fr\'echet mean of $\{\bm{u}^{\ast}_{ik}\}_{k=1}^N$ and $\{\bm{v}^{\ast}_{jk}\}_{k=1}^N$, respectively, and $\bar{\rho}_i$ and $\bar{\tau}_j$ as the arithmetic mean of $\{\rho_{ik}\}_{k=1}^N$ and $\{\tau_{jk}\}_{k=1}^N$, respectively. In this sense, the LP-size of $\bar{s}^{\LP}$ would be equal to one (see Result 1 in \hyperlink{link_Supplementary}{SUP}). In addition, we need to find mean frames. Calculating mean frame based on each frame vectors may violate the orthogonality condition. Thus the aim is to solve the following optimization problem $\forall{j}$,
\begin{mini}|s|
{\bm{n}_j,\bm{b}_j,\bm{b}_j^\perp\in{\mathbb{S}^2}}{\sum_{k=1}^{N}\left(d_g^2(\bm{n}_j,\bm{n}^{\ast}_{jk})+d_g^2(\bm{b}_j,\bm{b}^{\ast}_{jk})+d_g^2(\bm{b}_j^{\perp},\bm{b}^{\ast\perp}_{jk})\right)^\frac{1}{2},}
{}{}
\label{equ:Optimization}
\addConstraint{d_g(\bm{n}_j,\bm{b}_j)=d_g(\bm{b}_j,\bm{b}_j^\perp)=\frac{\pi}{2}}.
\end{mini}
Let $\hat{F}_j=(\bar{\bm{n}}_j,\bar{\bm{b}}_j,\bar{\bm{b}}_j^\perp)$ where $\bar{\bm{n}}_j$, $\bar{\bm{b}}_j$, and $\bar{\bm{b}}^\perp_j$ are Fr\'echet means of $\{\bm{n}^{\ast}_{jk}\}_{k=1}^N$, $\{\bm{b}^{\ast}_{jk}\}_{k=1}^N$, and $\{\bm{b}^{\ast\perp}_{jk}\}_{k=1}^N$ respectively. $\hat{F}$ may or may not belong to $SO(3)$ (i.e., $\hat{F}_j$ is not necessarily a frame). Thus $\hat{F}_j$ is the answer of the optimization \eqref{equ:Optimization} without its constrain. We rotate the initial frame $F=\tilde{I}$ to be as close as possible to $\hat{F}_j$ component wise (see \Cref{fig:mean_frame2}). Since rotation preserves the frame orthogonality, the aligned frame approximates the solution. For this purpose we use \Cref{Algorithm:frame_alignment} as a gradient descent approach. Note that the PNS mean estimates mean direction based on the euclideanized data, so we consider PNS mean instead of Fr\'echet mean. Also, we use the geometric mean instead of the arithmetic mean.\par
To reduce convergence time, choosing an appropriate initial frame for \Cref{Algorithm:frame_alignment} is essential. Analogous to GPA, we can consider the initial frame as $F=R(\frac{\bm{1}_3}{\|\bm{1}_3\|},\frac{\bar{\bm{n}}_j+\bar{\bm{b}}_j+\bar{\bm{b}}_j^\perp}{\|\bar{\bm{n}}_j+\bar{\bm{b}}_j+\bar{\bm{b}}_j^\perp\|})\tilde{I}$, i.e., $\tilde{I}$ is rotated such that its unit centroid coincides with the unit centroid of $\hat{F}_j$. Alternatively, since in practice $d_g(\bar{\bm{n}}_j,\bar{\bm{b}}_j)\approx\frac{\pi}{2}$, and based on the fact that we defined $\bm{b}_j^\perp$ as the cross product of $\bm{n}_j$ and $\bm{b}_j$, we accelerate the convergence by defining the initial frame based on $\bar{\bm{n}}_j$ and $\bar{\bm{b}}_j$. Assume $\bm{\mu}_j\in{\mathbb{S}^2}$ the middle point of the shortest geodesic connecting $\bar{\bm{n}}_j$ and $\bar{\bm{b}}_j$. We move in opposite directions from $\bm{\mu}$ toward $\bar{\bm{n}}_j$ and $\bar{\bm{b}}_j$ by angle $\frac{\pi}{4}$ to reach two points. We consider these points as $\bm{n}_j$ and $\bm{b}_j$ because, $d_g(\bm{n}_j,\bm{b}_j)=\frac{\pi}{2}$, and they have equal distance $\frac{|d_g(\bar{\bm{n}}_j,\bar{\bm{b}}_j)-\frac{\pi}{2}|}{2}$ to $\bar{\bm{n}}_j$ and $\bar{\bm{b}}_j$ (see \hyperlink{link_Supplementary}{SUP}).
\begin{algorithm}
\setstretch{1.1}
\caption{Frame alignment.}
\label{Algorithm:frame_alignment}
\begin{algorithmic}
\REQUIRE Unit vectors $\bar{\bm{n}}$, $\bar{\bm{b}}$, $\bar{\bm{b}}^\perp$, initial frame $F$ (e.g., $F=\tilde{I}$), step size $\delta$, and threshold $\varepsilon$
\ENSURE Aligned frame $F$ to $(\bar{\bm{n}}$,$\bar{\bm{b}}$,$\bar{\bm{b}}^\perp)$
\STATE $\Delta \leftarrow \left(d_g(F(1),\bar{\bm{n}})^2+d_g(F(2),\bar{\bm{b}})^2+d_g(F(3),\bar{\bm{b}}^\perp)^2\right)^\frac{1}{2}$
\WHILE{$\Delta \geq \varepsilon$}
\FOR{$i\leftarrow 1$ to 3}
\STATE $F \leftarrow$ $R(F(i),U(F(i)+\delta\cdot{T(F(i),\bm{n})}))F$
\ENDFOR
\STATE $\Delta \leftarrow \left(d_g(F(1),\bar{\bm{n}})^2+d_g(F(2),\bar{\bm{b}})^2+d_g(F(3),\bar{\bm{b}}^\perp)^2\right)^\frac{1}{2}$
\ENDWHILE
\end{algorithmic}
\footnotesize($T(\bm{v},\bm{w})$ is the tangent vector at $\bm{v}$ pointing toward $\bm{w}$, $U(\bm{v})=\frac{\bm{v}}{\|\bm{v}\|}$, and $R$ is the spherical rotation matrix.)
\end{algorithm}

\subsection{Convert LP-ds-rep to GP-ds-rep}
\Cref{sec:local_frames} and \Cref{sec:Local_parameterization} discuss how to obtain an LP-ds-rep from a GP-ds-rep. For several reasons, e.g., for visualization, we may need to reverse the procedure. For GP-ds-rep visualization, it is sufficient to draw spokes individually. To visualize an LP-ds-rep, we convert it to a GP-ds-rep. We start from $\tilde{I}$ as the s-centroid frame. Then, we reconstruct frames by finding the position and orientation of the frame's children based on $\tilde{I}$. Afterward, we find the information of grandchildren frames based on their parents and so on.\par
Let frame $F^{\ast}$ be in the coordinate system of its parent $F^\dagger$. To find $F^{\ast}$ based on GCS, we rotate $F^\dagger$ by $R_2R_1$ such that $R_2R_1F^\dagger=\tilde{I}$. Then $[R_2R_1]^{-1}F^{\ast}$ is the representation of $F^{\ast}$ in GCS. Similarly, we find the direction of connections and spokes in GCS.\par
Finding the mean shape of a set of objects' boundaries without an alignment is almost impossible. But we can use LP-ds-rep to estimate the mean boundary without alignment. First, we calculate the mean LP-ds-rep. Then, we convert the mean LP-ds-rep to a GP-ds-rep. Finally, we generate the implied boundary from the GP-ds-rep as demonstrated in \citep{liu2021fitting}. Therefore, it is possible to approximate the mean boundary without the alignment, which shows the power of LP-ds-rep.
\section{Hypothesis testing}\label{sec:hypothesis_testing}
Let $A=\{s_{Ai}\}_{i=1}^{N_1}$ and $B=\{s_{Bi}\}_{i=1}^{N_2}$ be two groups of either GP-ds-reps or LP-ds-reps of sizes $N_1$ and $N_2$. Let $K$ be the total number of GOPs. To test GOPs' mean difference, we design $K$ partial tests. Let $\bar{s}_{A}(k)$ and $\bar{s}_{B}(k)$ be the observed sample mean of the $k$th GOP from $A$ and $B$ respectively. The partial test is $H_{0k}:\bar{s}_{A}(k)=\bar{s}_{B}(k)$ versus $H_{1k}:\bar{s}_{A}(k)\neq\bar{s}_{B}(k)$. Note for GP-ds-rep $K=(n_p+2n_s+1)$, and for LP-ds-rep $K=(2n_s+5n_p+1)$. Also, for GP-ds-rep tests we pre-align the pooled group by GPA.\par
To test mean differences, we adapted a non-parametric permutation test with minimal assumptions similar to Styner's approach \citep{Styner2006}. For the univariate data i.e., vectors' lengths and shapes' sizes, the test statistic is t-statistic $T=\frac{\bar{x}-\bar{y}}{S_p\sqrt{\frac{1}{N_1}+\frac{1}{N_2}}}$ where $S_p$ is the pooled standard deviation. For the multivariate data i.e., euclideanized directions and GP-ds-rep skeletal positions, the test statistic is Hotelling's T$^2$ metric $T^2=(\bar{\bm{x}}-\bar{\bm{y}})^T\hat{\Sigma}^{-1}(\bar{\bm{x}}-\bar{\bm{y}})$, where $\hat{\Sigma}$ is an unbiased estimate of common covariance matrix \citep[ch.3]{mardia1982multivariate}. Given the pooled group $\{A,B\}$, the permutation method randomly partitions $B$ times the pooled group into two paired groups of sizes $N_1$ and $N_2$ without replacement, where usually we consider $B\geq{10^4}$. Afterward, it measures the test statistic between the paired groups. The empirical $p$-value for the $i$th GOP is $\eta_i=\frac{{1+\sum_{j=1}^{B}\chi_E(|T_{ij}|\geq{T_{io}})}}{B+1}$, where $T_{io}$ is the $i$th observed test statistics, $T_{ij}$ is the $j$th permutation test statistic, and $\chi_E$ is the indicator function i.e., $\chi_E(\varphi)=1$ if $\varphi$ is true, otherwise $\chi_E(\varphi)=0$.\par
In order to account for the problem of multiple hypothesis testing, one could use Bonferroni's method \citep{bonferroni1936teoria}. Bonferroni's method tests each hypothesis at level $\alpha/K$ and guarantees the probability of at least one type I error $P(v\geq{1})$ be less than significance level ${\alpha}$. Since Bonferroni's method is highly conservative we prefer to apply \textit{Benjamini-Hochberg} (BH) \citep{benjamini1995controlling} as a more moderate approach. 

\section{Evaluation}\label{sec:eval}

\subsection{Data}\label{sec:data}
To test our method, we study the hippocampal difference between early PD and CG at baseline. Data are provided by ParkWest (\url{http://parkvest.no}), in cooperation with Stavanger University Hospital (\url{https://helse-stavanger.no}). At the baseline, we have 182 \textit{magnetic resonance} (MR) images for PD and 108 for CG with corresponding segmentation of hippocampi. As described in \Cref{sec:s-rep},  GP-ds-reps are fitted to left hippocampi by SlicerSALT (\url{http://salt.slicer.org}) and re-parametrize into LP-ds-reps. 
For the model fitting, we used GP-ds-reps with 122 spokes consisting of 51 up, 51 down, and 20 crest spokes. As up and down spokes share the same tail position, we have in total 71 tail positions. Thus, for LP-ds-rep, we have 122 spokes, 71 local frames, and 71 connections. Before analyzing the Parkinson data, we first study our method based on simulations.

\subsection{Simulation}\label{sec:simulation}
\begin{figure}[ht]
\centering
  \boxed{\includegraphics[width=0.7\textwidth]{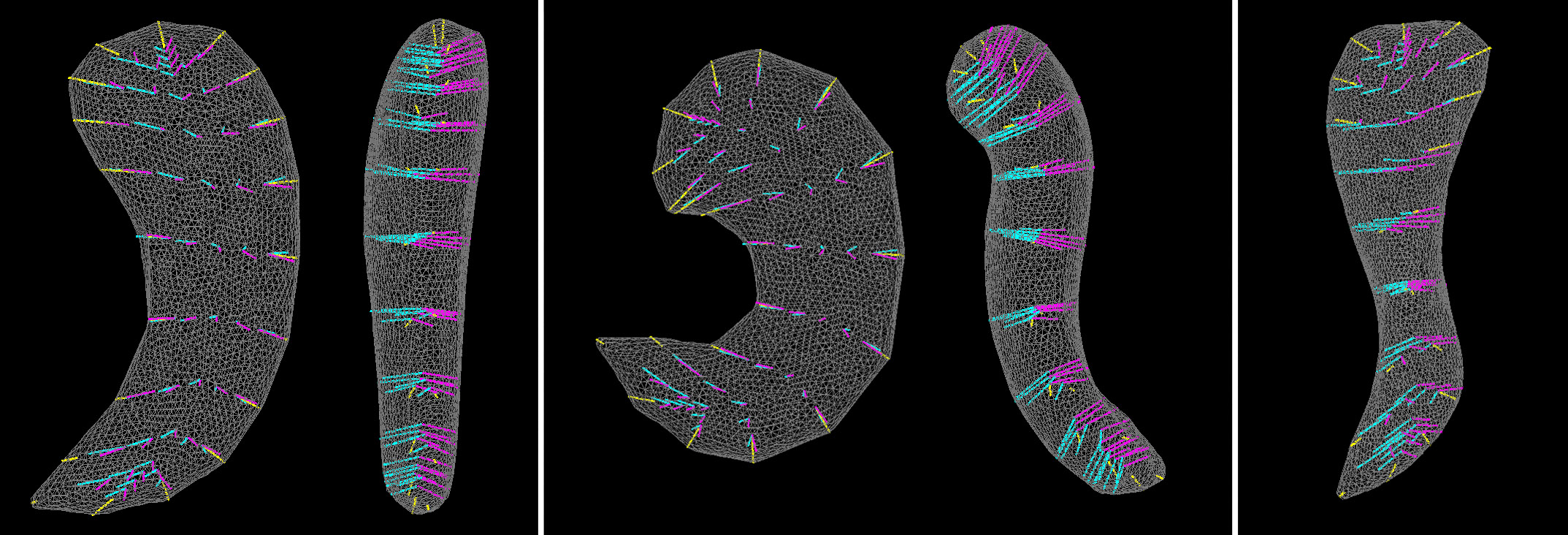}}
\caption{\centering\footnotesize{Skeletal deformation by LP-ds-rep. Left: A ds-rep with its implied boundary in two angles. Middle: Shape bending by spinal frame rotation about $\bm{n}$ and $\bm{b}^\perp$ axes. Right: Shape twisting by spinal frames rotation about $\bm{b}$ axis.}}
\label{fig:shape_deformation}
\end{figure}

In statistical shape analysis generating random shapes is a matter of interest. Designing simulation based on GP-ds-rep is challenging as we usually need to identify a local frame to bend or twist the object locally. It turned out that LP-ds-rep support naturally skeletal deformations. We can stretch, shrink, bend, and twist the skeletal by manipulating the frames' orientations and vectors' lengths. Then, we convert LP-ds-rep to GP-ds-rep to generate the boundary. Consequently, we can add variation to a set of deformed LP-ds-reps' GOPs to simulate random ds-reps. \Cref{fig:shape_deformation} shows a deformed hippocampus including bending and twisting. The deformation is done by rotating six spinal frames. \par
\begin{figure}[ht]
\centering
  \boxed{\includegraphics[width=0.7\textwidth]{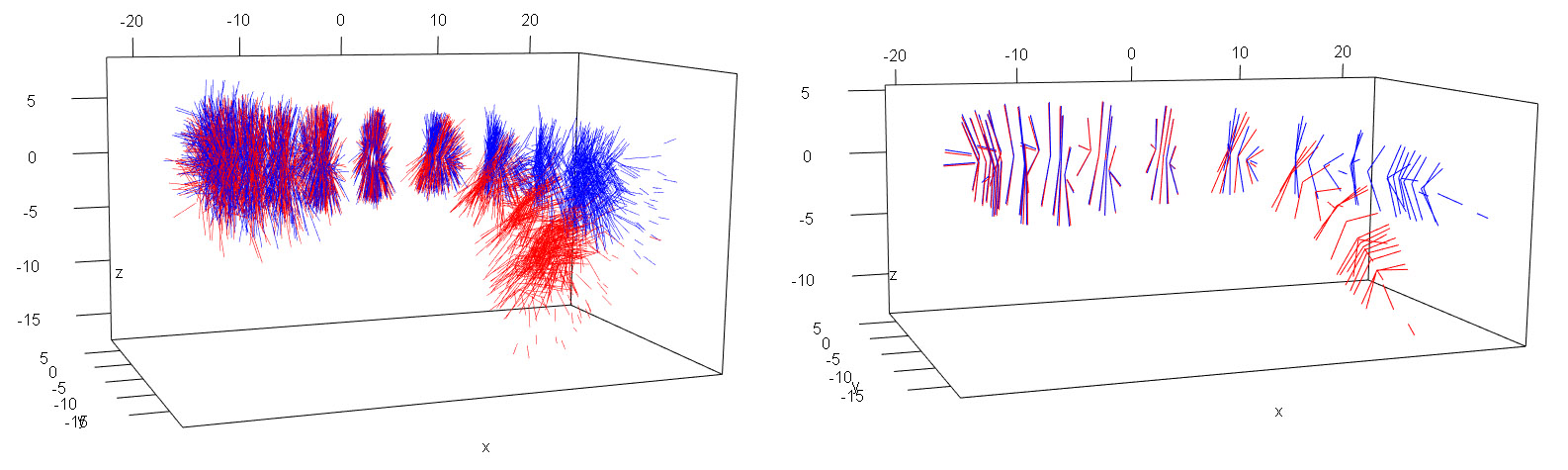}}
\caption{\centering\footnotesize{ds-rep simulation by LP-ds-rep. Left: Blue and red indicate twenty samples of two groups of simulated ds-reps. Right: Overlaid mean LP-ds-reps.}}
\label{fig:simulation}
\end{figure}

For the simulation study, we select the mean LP-ds-rep of CG from the ParkWest data as a template. Based on the template, we generate two LP-ds-rep groups of sizes 150 with different amount of tail bending, i.e. bending in local region. Such bending was observed for example in \citep{pizer2003object} between schizophrenics and controls. Let $M_p(\bm{\mu},\kappa)$ denotes von Mises-Fisher distribution with mean $\bm{\mu}$ and concentration parameter $\kappa$ on $\mathbb{S}^{p-1}$  \citep{dhillon2003modeling}. For the special case $p=2$ we assume the distribution in radian i.e., $\theta,\mu\in[0,2\pi)$ if $\theta\sim{M_2(\mu,\kappa)}$. Given random rotation angle of bending $\theta\sim{M_2(\mu=0,\kappa=100)}$ for the first group and $\theta\sim{M_2(\mu=\frac{-\pi}{15},\kappa=100)}$ for second group, we simulate the orientation of three spinal frames by successively rotating them about their $\bm{b}^\perp$-axis with $[R_2R_1]^{-1}R(\bm{e}_3,(\cos{\theta},0,\sin{\theta})^T)\tilde{I}$. This means, the tails in the second group is in average successively bend $12^\circ$ downward for three consecutive spinal frames. Chosen frames are the closest ones on the hippocampus tail to the s-centroid. Thus, in total, we have a slight downward bending about $36^\circ$ at the hippocampus tail.
Finally, by preserving frame orthogonality, we add noise to all directions by $M_3(\bm{\mu},\kappa)$, where $\kappa$ for frames' vectors, spokes, and connections is equal to $600$, $250$, and $5000$, respectively. Further we added noise to vectors' lengths by the truncated normal distribution $\psi(\mu,\sigma,a>0,b<\infty)$ where $\mu$ is the vector length of the template, and parameters $\sigma$, $a$, and $b$ are heuristically chosen. As a result, we have two groups of random LP-ds-reps, which are approximately similar in most of their GOPs but only different in the orientation of three frames. \Cref{fig:simulation} illustrates twenty samples of each group in blue and red. Note that LP-ds-reps are not aligned, but since we reconstruct them from the s-centroid frame, shapes have Bookstein's alignment \citep[Ch. 2]{dryden2016statistical} because the s-centroid frames are perfectly aligned.\par
\begin{figure}[ht]
     \centering
     \begin{subfigure}[b]{0.32\textwidth}
         \centering
         \includegraphics[width=\textwidth]{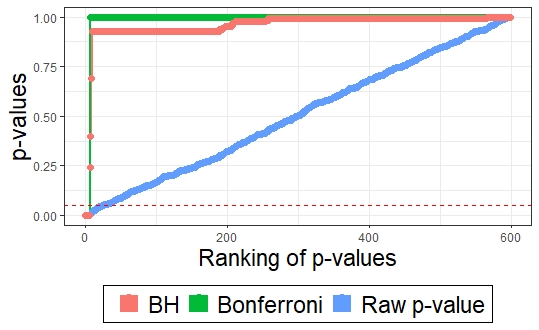}
         \caption{\centering\footnotesize{LP-ds-rep $p$-values}}
         \label{fig:p-value-LP1}
     \end{subfigure}
     \hfill
     \begin{subfigure}[b]{0.32\textwidth}
         \centering
         \includegraphics[width=\textwidth]{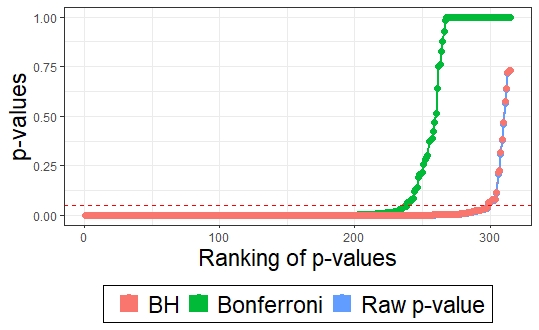}
         \caption{\centering\footnotesize{GP-ds-rep $p$-values}}
         \label{fig:p-value-GP1}
     \end{subfigure}
     \hfill
     \begin{subfigure}[b]{0.32\textwidth}
         \centering
         \includegraphics[width=\textwidth]{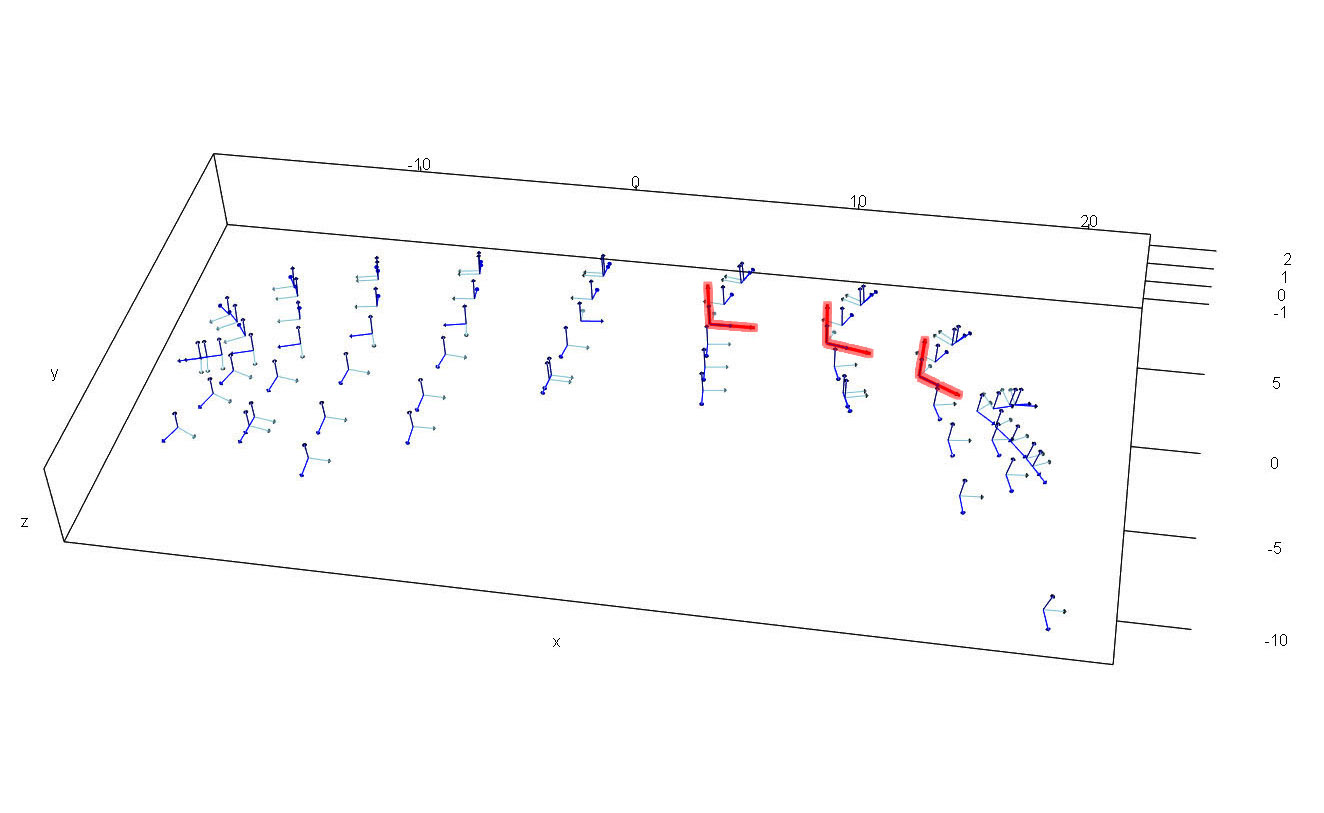}
         \caption{\centering\footnotesize{Significant frames}}
         \label{fig:sigFrames}
     \end{subfigure}
        \caption{\centering\footnotesize{LP-ds-rep vs. GP-ds-rep. (a,b) Raw and adjusted $p$-values by BH and Bonferroni. The dotted line indicates significance level $\alpha=0.05$. (c) Blue arrows indicate local frames. Red indicates statistically significant frame directions after the BH adjustment using LP-ds-rep analysis.}}
        \label{fig:LP_vs_GP_pvalues}
\end{figure}

Hypothesis test on LP-ds-rep from \Cref{sec:hypothesis_testing} correctly detects significant frame directions and label almost all other GOPs as statistically non-significant given a significance level $\alpha=0.05$. On the contrary, the test on GP-ds-rep indicates a large number of false positives, i.e., almost all of the positions and directions are statistically significant (see \Cref{fig:LP_vs_GP_pvalues}). This example confirms our observation from \Cref{fig:arms_pdm} in \Cref{sec:introduction} and highlights the fact that GP-ds-rep analysis could be extremely biased.
\subsection{Real data analysis}\label{sec:results} The Parkinson data set described in \Cref{sec:data} was studied earlier by \citep{apostolova2012hippocampal}  based on radial distance analysis and parallel slicing and showed some regional atrophy. Since shape correspondence in parallel slicing is controversial, we attempt to reanalyze data by utilizing LP-ds-rep.\par
First let us compare the shape sizes, see \Cref{table:LP_GP_sizes}. Tests on shape size indicate no significant difference. However, volume measurement confirms the LP-size is more compatible with the object volume as for both, the mean object volume and the LP-size of CG are greater than PD. In opposite the mean GP-size of CG is smaller than PD.\par
\begin{figure}[ht]
     \centering
     \begin{subfigure}[b]{0.32\textwidth}
         \centering
         \includegraphics[width=\textwidth]{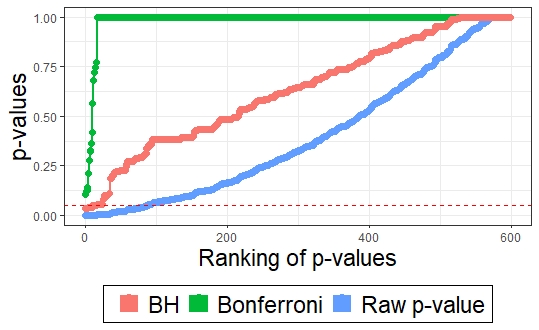}
         \caption{\centering\footnotesize{LP-ds-rep $p$-values}}
         \label{fig:p-value-LP}
     \end{subfigure}
     \hfill
     \begin{subfigure}[b]{0.32\textwidth}
         \centering
         \includegraphics[width=\textwidth]{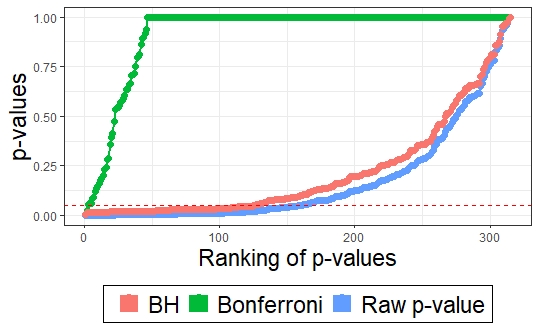}
         \caption{\centering\footnotesize{GP-ds-rep $p$-values}}
         \label{fig:p-value-GP}
     \end{subfigure}
     \hfill
     \begin{subfigure}[b]{0.32\textwidth}
         \centering
         \includegraphics[width=\textwidth]{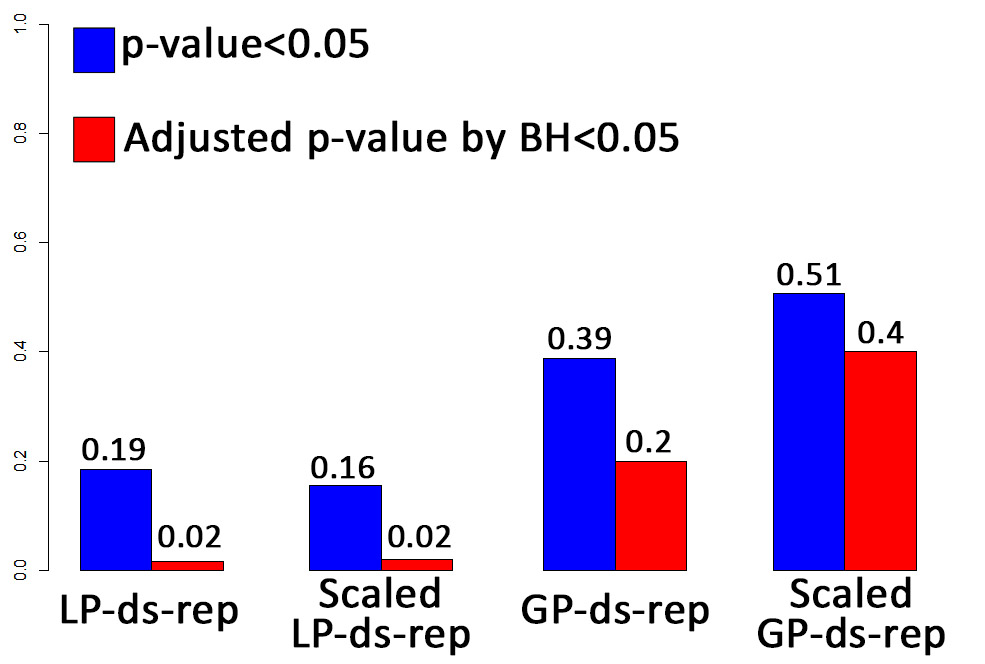}
         \caption{\centering\footnotesize{Scaling effect}}
         \label{fig:p-value-scaling}
     \end{subfigure}
        \caption{\centering\footnotesize{Test on real data. (a,b) Raw and adjusted $p$-values by BH and Bonferroni. The dotted line indicates significance level 0.05. (c) Bar plot shows the scaling effect on the percentage of $p\textup{-values}<0.05$.}}
        \label{fig:pvalues_shape_vs_sizeAndShape}
\end{figure}

\begin{table}
    \small 
    \centering
    \begin{tabular}{|c|c|c|c|c|c|}
    \hline
           & Mean CG & Mean PD & SD CG & SD PD & p-value \\\hline
         GP-size of spokes' tips & 161.05 & 162.51 & 8.97 & 8.62 & 0.17 \\\hline
         Object volume (mm$^3$) & 3352.23 & 3271.44 & 563.39 & 616.68 & 0.26 \\\hline
         LP-size & 536.31 & 527.60 & 36.09 & 38.86 & 0.06 \\\hline
    \end{tabular}
    \caption{\centering\footnotesize{T-test on shape size.}}
    \label{table:LP_GP_sizes}
\end{table}
\begin{figure}[ht]
\centering
  \boxed{\includegraphics[width=0.97\textwidth]{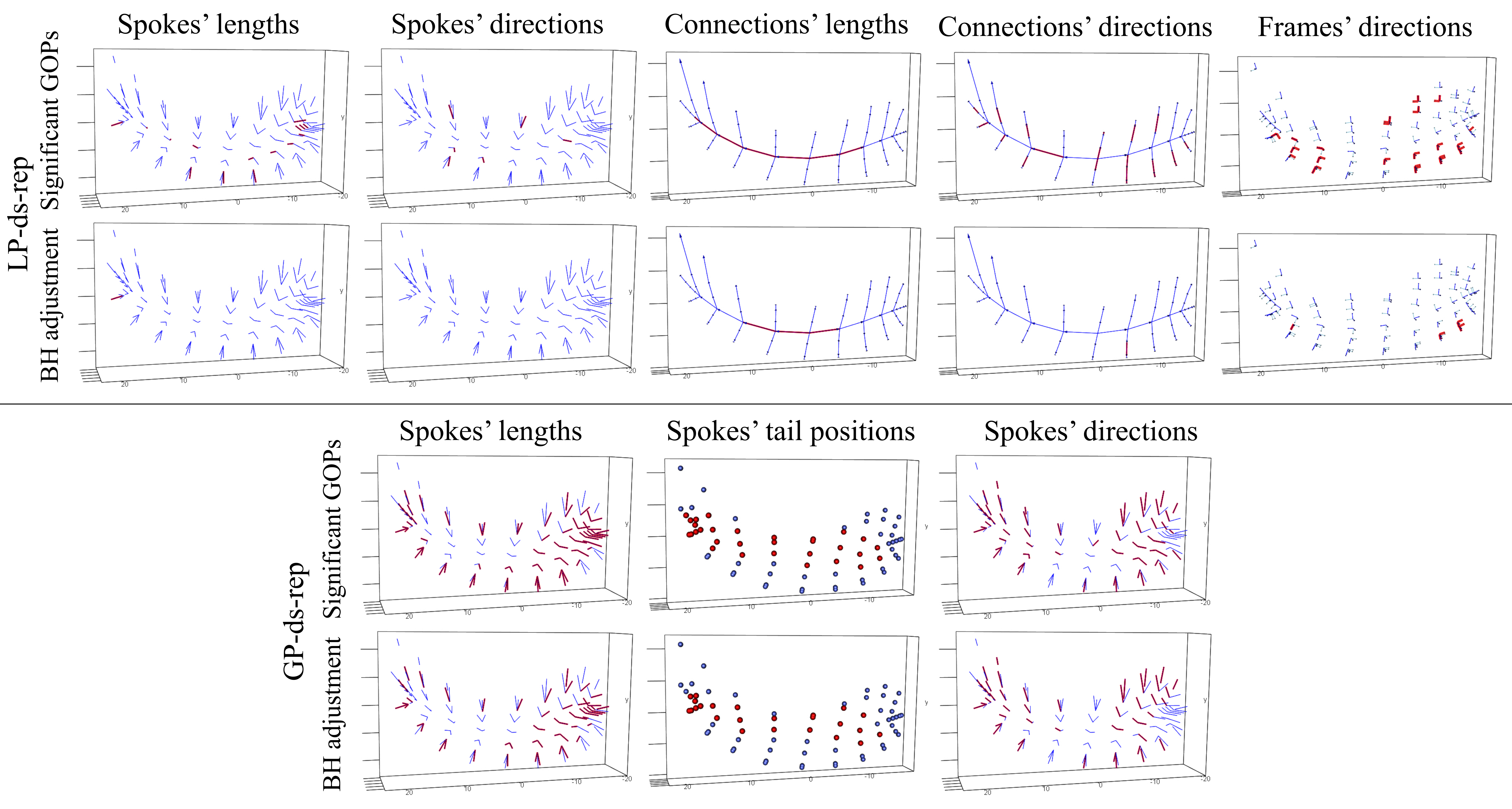}}
\caption{\centering\footnotesize{ds-rep significant GOPs. Red indicate significant GOPs. FDR=0.05 for BH adjustment.}}
\label{fig:result_test}
\end{figure}
\Cref{fig:result_test} illustrates significant LP-ds-rep and GP-ds-rep GOPs before and after BH adjustment in red. In LP-ds-rep, all the spokes directions are insignificant. In contrast, about 40\% of GP-ds-rep spokes' directions are significant. Also, in LP-ds-rep, there are a few significant connection and frame directions after the adjustment. Based on the LP-ds-rep analysis, it seems the main difference comes from connections' length on the spine. \Cref{fig:p-value-LP} and \Cref{fig:p-value-GP} show sorted $p$-values before and after adjustment. Based on Bonferroni adjustment, PD and CG are similar because almost all adjusted $p$-values are greater than 0.05. But based on raw and BH $p$-values, about half of the GP-ds-rep GOPs are significant, while in LP-ds-rep, we have only a few.\par
In addition, we analyzed the shapes without scaling. Detailed results are available in \hyperlink{link_Supplementary}{SUP}. The general belief is scaling makes shapes more similar. But \Cref{fig:p-value-scaling} expresses the percentage of significant GP-ds-rep GOPs increases dramatically after the scaling. In other words, scaling increases the number of raw $p$-values less than the level of significance $\alpha=0.05$ and consequently increases the number of BH adjusted $p$-values less than FDR=0.05. A possible explanation is that GPA tries to make shapes as close as possible by reducing GOPs' variation. By removing the scale, GPA reduces the variation even more. Hotelling's T$^2$ metric is proportional to the inverse common covariance matrix. So by reducing the variation, the test statistic increases, and consequently, the $p$-value decreases. On the contrary, the LP-ds-rep is not sensitive to scaling as it is alignment-independent.
\section{Conclusion}\label{sec:discussion}
Generally, it is common to detect locational dissimilarity between two groups of objects based on the alignment. We showed that alignment-dependent analysis such as ds-rep analysis in GCS could be highly biased. We described GP-ds-rep as a conventional ds-rep parameterization in GCS. We introduced LP-ds-rep as a novel parameterization with a hierarchical structure based on locally fitted frames to overcome inherent challenges caused by alignment. We have defined a mean LP-ds-rep and discussed object deformation and simulation with LP-ds-rep. We explained how to estimate boundary mean shape without alignment. We compared LP-ds-rep with GP-ds-rep to show the advantages of LP-ds-rep. For comparison, we applied simulation and real data analysis. The simulation confirmed that the hypothesis test based on GP-ds-rep for two groups of ds-rep with a slight difference results in many false positives while LP-ds-rep indeed detected the differences. For the real data, we studied left hippocampi of early PD vs. CG. Although hypothesis tests on GP-ds-rep indicated many significant GOPs, tests on LP-ds-rep showed only a few, which seems medically more reasonable. Also, data analysis exposed GP-ds-rep sensitivity to the scaling. We concluded that PD and CG groups are very similar, but the main difference comes from the spinal stretch of the skeletal sheet.
\section*{Acknowledgments}
This research is funded by the Department of Mathematics and Physics of the University of Stavanger (UiS). Special thanks to Profs. Stephen M. Pizer (UNC), Steve Maron (UNC), James Damon (UNC), and Jan Terje Kvaløy (UiS) for insightful discussions and inspiration for this work. We are indebted to Prof. Guido Alves (UiS) for providing ParkWest data. We also thank Zhiyuan Liu (UNC) for the model fitting toolbox.
\bigskip
\begin{center}
{\large\bf SUPPLEMENTARY MATERIALS}
\end{center}
\textbf{Supplementary:} \hypertarget{link_Supplementary} SUP materials referenced in this work are available in this pdf. (pdf)
\textbf{R-code:} In Supplementary.zip, simulation codes and files are placed. (zip)
\bibliographystyle{spbasic}
\bibliography{myBibFile}

\end{document}